\documentclass[fleqn,usenatbib]{mnras}

\usepackage[T1]{fontenc}
\usepackage{ae,aecompl}

\usepackage{graphicx}	
\usepackage{amsmath}	
\usepackage{amssymb}	
\usepackage{multirow}
\usepackage{booktabs}
\usepackage[none]{hyphenat}
\usepackage{subcaption}
  \newcommand{\lbol}{$L_{\rm bol}$}
  \newcommand{\ledd}{$L_{\rm Edd}$}
  \newcommand{\gro}{GRO J1655-40}
  \newcommand{\gx}{GX 339-4}
  \newcommand{\lx}{$L_{\rm X}$}
  \newcommand{\mbh}{$M_\mathrm{BH}$}

  \newcommand{\mdot}{$\dot{m}$}
  \newcommand{\msun}{M$_{\sun}$}
  \newcommand{\nbmc}{$N_\mathrm{BMC}$}
  \newcommand{\nbmcr}{$N_\mathrm{BMC,r}$}
  \newcommand{\nbmct}{$N_\mathrm{BMC,t}$}
  \newcommand{\xmm}{{\it XMM-Newton}}
  \newcommand{\xte}{XTE J1550-564}

\title[Are NLS1s highly accreting low-$M_\mathrm{BH}$ AGNs?]{Are narrow-line Seyfert 1 galaxies highly accreting low-$M_\mathrm{BH}$ AGNs?}

\author[J. K. Williams et al.]{
James K. Williams,$^{1}$\thanks{E-mail: jwilli91@masonlive.gmu.edu}
Mario Gliozzi,$^{1}$
Ross V. Rudzinsky$^{2}$
\\
$^{1}$ Department of Physics and Astronomy,
George Mason University, 4400 University Drive, Fairfax, VA 22030\\
$^{2}$ Department of Physics, University of California, 366 LeConte Hall, Berkeley, CA 94720
}

\date{Accepted 2018 July 7. Received 2018 July 5; in original form 2018 April 5}

\pubyear{2017}

\begin{document}
\label{firstpage}
\pagerange{\pageref{firstpage}--\pageref{lastpage}}
\maketitle

\begin{abstract}
In this work, we test the hypothesis that narrow-line Seyfert 1 galaxies (NSL1s) are active galactic nuclei in their early phase and are therefore younger and more active than the more common broad-line Seyfert 1 galaxies (BLS1s). If that is true, then NLS1s should, on average, have lower black hole (BH) masses and higher accretion rates than BLS1s. To test this, we use a sample of 35 NLS1s and 54 BLS1s with similar X-ray luminosity distributions and good \xmm\ observations. To determine the BH mass \mbh, we apply an X-ray scaling method that is independent of any assumptions on the broad-line region dynamics and the inclination of the objects. We find that, on average, NLS1s have lower BH masses, but the difference between the average \mbh\ of NLS1s and BLS1s in our sample is only marginally significant (at the 2.6 sigma level). According to a Kolmogorov--Smirnov test, the distribution of \mbh\ values of NLS1s is different from that of BLS1s at the 99\% confidence level. Even stronger differences between NLS1s and BLS1s are inferred when the accretion rate distributions of NLS1s are compared to BLS1s, suggesting that the two populations are indeed distinct. Our study also indicates that the \mbh\ values (both for NLS1s and BLS1s) determined with the X-ray scaling method are fully consistent with those obtained using reverberation mapping.
\end{abstract}

\begin{keywords}
Galaxies: active -- Galaxies: nuclei -- X-rays: galaxies
\end{keywords}



\section{Introduction}

Narrow-line Seyfert 1 galaxies (NLS1s) are distinguished from the more common broad-line Seyfert 1 galaxies (BLS1s) by the properties of their emission lines. In the optical spectrum, NLS1s have an [O\,\textsc{iii}]/H\,$\beta$ ratio of less than 3 and a full width at half maximum (FWHM) H\,$\beta$ of less than 2000 km s$^{-1}$. In the X-ray spectrum, NLS1s tend to have steep spectra and strong variability. These differences have led to the hypothesis that NLS1s are active galactic nuclei (AGNs) in their early phase and therefore are younger and more active \citep[e.g.,][]{boro92,grup04}. If NLS1s are younger and more active, they should, on average, have lower black hole (BH) masses and higher accretion rates than BLS1s.

Several optical estimates of the black hole mass \mbh\ of NLS1s suggested that NLS1s lie below the \mbh--$\sigma_\mathrm{bulge}$ curve ($\sigma_\mathrm{bulge}$ is the velocity dispersion of the galaxy's bulge), which implies that they are accreting at higher rates than BLS1s \citep[e.g.,][]{grup04,math05a,math05b}. However, other research challenged the use of optical methods for determining \mbh\ in NLS1s on the basis that highly accreting objects produce high radiation pressure that partially counteracts the action of the BH gravitational pull, leading to values of \mbh\ that are systematically underestimated \citep[e.g.,][]{marc08}. Another possibility is that NLS1s are simply ordinary Seyfert 1s with a disk-shaped broad-line region (BLR) and a line of sight that is pole-on; in that case, the difference between NLS1s and BLS1s disappears \citep[e.g.,][]{decr08}. Furthermore, past studies suggest that NLS1s may not be a homogeneous class. For example, optically selected and X-ray-selected NLS1s appear to have different properties \citep[e.g.,][]{grup04a}. In addition, recent results from \textit{Fermi LAT} suggest that a small subset of NLS1s are likely to be jet-dominated \citep[e.g.,][]{fosc15}.

To test the hypothesis introduced above, it is crucial to make accurate measurements of \mbh\ and accretion rate $\dot{m}$. The best and most direct way of obtaining \mbh\ is by dynamical methods. For our own Galaxy, we can derive \mbh\ by directly observing the motion of gas and stars near the black hole, but this requires that the region of influence of the black hole can be resolved by our telescope. Therefore, this method works only for galaxies that are nearby and that have weakly active nuclei. For AGNs farther away, the best method we have is reverberation mapping, which derives \mbh\ by measuring the size and kinematics of the BLR by observing the light-speed time delay $\tau$ as changes in the continuum emission propagate out from the central accretion disk and cause changes in lines emitted by structures farther away from the central BH \citep[e.g.,][]{pete06}.

However, these direct methods have some limitations when applied to our particular problem. Applying the most direct method for AGNs, reverberation mapping, to highly accreting objects can be questionable when the dynamics of the BLR may be affected by radiation pressure and not just \mbh\ gravitational force, or with small viewing angles, or both. For NLS1s (and highly accreting objects in general), it is important to use methods to constrain \mbh\ that are independent of any assumptions on the BLR dynamics and geometry. Therefore, for this work we used an X-ray scaling method of determining \mbh\ \citep{shap09}. The X-ray scaling method does not depend on any assumptions about the BLR, nor does it need to make a correction for radiation pressure.

It is now generally accepted that hard X-rays (> 2 keV) are produced primarily by inverse Comptonization, where seed photons, originally emitted in the optical and UV directly from the accretion disk, are upscattered several times and become hard X-rays in the corona. The energy band of the original photons depends heavily on \mbh\ and \mdot. This process appears to be ubiquitous in all BH systems, which is a major motivation for using X-rays to measure \mbh. It means we can use the X-ray scaling method for all types of black holes. Indeed, this scaling method, originally developed for Galactic black holes \citep{shap09} was later successfully extended to supermassive black holes  \citep{glio11} and also to ultraluminous X-ray sources, which may host intermediate-mass BHs \citep{jang18}. In the X-ray scaling method, the value of \mbh\ for any black hole system, accreting at a moderate or a high level, can be obtained by scaling the \mbh\ of a reference source, which is a stellar mass BH in a binary system whose parameters are tightly constrained via direct methods. The scaling factor is determined by the ratio of the normalization parameter in the Comptonization model used to fit the X-ray spectrum of the target to the analogous parameter of the reference source. Further details of this method are provided in Section 4. In principle, the method can be used for any BH for which most of the X-rays are produced by Comptonization, although care must be taken to account for absorption and reflection.

In this work, we applied the X-ray scaling method to a subsample of NLS1s and BLS1s drawn from the flux-limited sample compiled by Zhou and Zhang (2010) using \xmm\ observations. Our goal was to investigate whether the distribution of \mbh\ and $\dot{m}$ of NLS1s is statistically different from that of BLS1s.  In Section 2, we briefly describe the sample and the data reduction. Section 3 reports the results of our X-ray spectral analysis. In Section 4, we show how we derived the BH masses and accretion rates using the X-ray scaling method and compare their respective distributions for NLS1s and BLS1s. Finally, in Section 5, we summarize our main findings and discuss their implications.

Throughout the paper, we use a cosmology based on a $\Lambda$CDM model with $H_0 = 71$ km s$^{-1}$ Mpc$^{-1}$, $\Omega_m = 0.27$, and $\Omega_v = 0.73$.

\section{Data Reduction}
We took our NLS1 and BLS1 samples from the flux-limited sample ($f_\mathrm{\,2-10\, keV} \geq 1\times10^{-12}\ \mathrm{ erg}\, \mathrm{s}^{-1} \mathrm{cm}^{-2}$) of Zhou and Zhang (2010), which contains 114 radio-quiet type I AGNs with hard X-ray (2\textendash10 keV) luminosities from $10^{41}$ to $10^{45}$ erg s$^{-1}$. Of the 114 AGNs in their sample, 16 are narrow emission line galaxies (NELGs), and those were not processed or used in this study. That left a total of 98, which we divided into a sample of 36 NLS1s and a second sample of 62 BLS1s. The two samples, which are constructed based on the similarity of their X-ray luminosity distributions, contain the brightest members of BLS1s and NLS1s observed by \xmm, and hence may represent the brightest tails of the two populations rather than the whole populations. Therefore, we acknowledge that our samples are starting points and not perfectly random. All 98 AGNs were processed with the \xmm\ Science Analysis System (SAS) version 15.0.0, and for each source, ancillary response (\textsc{arf}) and redistribution matrix (\textsc{rmf}) files were created. For sources with more than one \xmm\ observation, we chose the most recent one with a duration of at least 10 ks, using the EPIC cameras \citep[e.g.,][]{stru01,turn01}. Source photons were extracted in a circle with a radius of 30\arcsec\ centered on the source from the EPIC pn image, and the background was selected from a source-free region on the same CCD as the source and with a radius of 60\arcsec. Spectra were rebinned with \textsc{grppha} to have at least 20 counts per bin, although for some faint sources we used 15 instead. The recorded events were screened to remove known hot pixels and data affected by background flares, as recommended by the SAS user guide.

To assess the presence of pile-up, for each source we verified the EPIC camera observing mode and compared the background-subtracted source count rate with the appropriate threshold count rates above which sources are expected to suffer from pile-up according to the \textit{XMM-Newton Users Handbook}. Of the 89 AGNs for which we computed \mbh, 84 have count rates well below the pile-up thresholds. For the remaining five sources, following the SAS user guide, we quantified the incidence of pile-up using the SAS task \textsc{epatplot}. This tool provides plots of the distributions of single and double patterns, which are very sensitive to pile-up effects: deviations from the model curves indicate the presence of pile-up. Of the five bright AGNs, only Mrk 1383 and Mrk 841 show clear deviations from the theoretical single and double patterns curves, indicating that these two sources are indeed affected by pile-up (the level of pile-up was estimated using \textsc{webpimms} and it is of the order of 7\% for Mrk 1383 and 10\% for Mrk 841). Following the procedure recommended by the SAS user guide to mitigate the effects of pile-up on EPIC camera data, we excised the inner part of the extraction region using an annulus with inner and outer radii of 10\arcsec and 40\arcsec, respectively. After checking with \textsc{epatplot} that the new extraction regions yielded the correct single and double patterns, we extracted the spectra and ran the \textsc{arfgen} task again to account for the flux loss. A comparison of the spectra of Mrk 1383 and Mrk 841 using annular extraction regions with those using circular regions shows that the latter are fitted by flatter photon indices, as expected for piled-up spectra.

\section{X-ray Spectral Analysis}

All the AGNs were fitted using \textsc{xspec} \citep{arna96}. Since the X-ray scaling method depends only on the Comptonization component, we restricted the spectral fitting range to 2--10 keV. This also reduces possible spectral-fitting complications related to the characterization of the soft X-ray component. Our basic \textsc{xspec} model was wabs*bmc. WABS models both Galactic and intrinsic absorption. BMC is the bulk motion Comptonization model \citep{tita97}. When needed, Gaussian lines were added, most commonly to model prominent Fe K\,$\alpha$ emission lines near 6.4 keV.

BMC has the advantage of being generic enough to describe both thermal Comptonization and bulk motion Comptonization equally well. The model convolves the seed photons and a generic Comptonization Green's function giving a power law, which usually produces a good fit.

The BMC model has the following four parameters: (1) $kT$, the temperature of the thermal seed photons, (2) $\alpha$, the energy spectral index, (3) log $A$, the logarithm of the $A$ parameter, and (4) \nbmc, the BMC normalization. The spectral index $\alpha$ is related to the photon index $\Gamma$ by $\Gamma = 1 + \alpha$. The $A$ parameter is related to the Comptonization fraction $f$ by $f = A/(1+A)$. \nbmc\ is a function of luminosity $L$ and distance $d$ and is proportional to $L/d^2$. In our analysis, we started with reasonable guesses for these parameters and let them vary.

\begin{figure}
	\includegraphics[width=0.65\columnwidth, angle=-90]{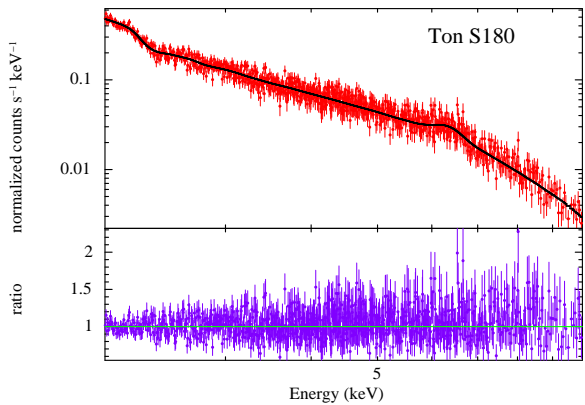}
	\caption{Best fit for the model wabs*(bmc+zgauss) for Ton S180. The lower panel shows the residuals.}
	\label{fig:tons180}
\end{figure}

To give just one illustration of the model, Figure~\ref{fig:tons180} shows a plot of the NLS1 Ton S180 spectrum fitted with the model wabs*(bmc+zgauss), which yielded the following best-fit parameters: $kT = 0.34_{-0.05}^{+0.04}$, $\alpha = 1.05_{-0.04}^{+0.07}$, log $A = 0.29$, and \nbmc\ $ = 5.3_{-0.2}^{+0.4}\times 10^{-5}$, with a reduced $\chi^2$ of 1.00 for 889 degrees of freedom (dof).
 
The results of our systematic and homogeneous spectral analysis of the NLS1 and BLS1 samples are reported in Tables \ref{tab:spectralnls1} and \ref{tab:spectralbls1_part1}, which show the best-fit values for 89 AGNs. For the other nine AGNs, the X-ray spectrum appears to be dominated by reflection or absorption or both, which hampers the proper characterization of the primary X-ray component, which is a requirement for applying the scaling method (see the next section for further details). All AGNs were reasonably well fitted with our baseline model, the reduced $\chi^2$ ranging from 0.63 to 1.37 with an average of 0.99. For NLS1s, $\alpha$ ranges from 0.54 to 1.44 with a mean of 1.00, whereas for BLS1s, $\alpha$ ranges from 0.23 to 1.14 with a mean of 0.67, confirming that NLS1s have on average steeper X-ray slopes. On the other hand, for the other three parameters, $kT$, log $A$, and \nbmc, there are no significant differences between NLS1s and BLS1s. For NLS1s, $kT$ ranges from 0.08 to 0.61 with a mean of 0.30, whereas for BLS1s, $kT$ ranges from 0.05 to 0.56 with a mean of 0.32. For NLS1s, log $A$ ranges from -0.45 to 2.00 with a mean of 0.60, whereas for BLS1s, log $A$ ranges from -1.17 to 2.00 with a mean of 0.55. For NLS1s, \nbmc\ ranges from $1.08\times 10^{-5}$ to $1.10\times 10^{-3}$ with a mean of $1.28\times 10^{-4}$, whereas for BLS1s, \nbmc\ ranges from $9.1 \times 10^{-6}$ to $9.93 \times 10^{-4}$ with a mean of $1.81 \times 10^{-4}$.

\begin{table*}
		\caption{NLS1 spectral data}
		
	\begin{center}
		\begin{tabular}{lccccc} 
			\toprule
			\toprule       
			Name & Model &  $kT$ &  $\alpha$ &  log $A$ & Norm \\
			(1) & (2) & (3) & (4) & (5) & (6) \\
			\midrule
			Mrk 335 & wabs*(bmc+zgauss) & $0.28_{-0.02}^{+0.02}$ & $0.90_{-0.05}^{+0.05}$ & -0.44 & $1.9_{-0.1}^{+0.1}\times 10^{-4}$ \\
			\noalign{\smallskip}
			I Zw 1 & wabs*(bmc+zgauss) & $0.20_{-0.20}^{+0.04}$ & $1.25_{-0.06}^{+0.05}$ & 0.17 & $1.6_{-0.2}^{+2.2}\times 10^{-4}$ \\
			\noalign{\smallskip}
			Ton S180 & wabs*(bmc+zgauss) & $0.34_{-0.05}^{+0.04}$ & $1.05_{-0.04}^{+0.07}$ & 0.29 & $5.3_{-0.2}^{+0.4}\times 10^{-5}$ \\
			\noalign{\smallskip}
			Mrk 359 & wabs*(bmc+zgauss) & $0.26_{-0.03}^{+0.03}$ & $0.63_{-0.08}^{+0.08}$ & 0.15 & $8.6_{-0.2}^{+0.2}\times 10^{-5}$ \\
			\noalign{\smallskip}
			Mrk 1014 & wabs*bmc & $0.30$ & $1.02_{-0.19}^{+0.21}$ & 0.34 & $1.1_{-0.9}^{+0.1}\times 10^{-5}$ \\
			\noalign{\smallskip}
			Mrk 586 & wabs*bmc & $0.19$ & $1.21_{-0.08}^{+0.08}$ & 1.22 & $2.0_{-0.2}^{+0.2} \times 10^{-5}$ \\
			\noalign{\smallskip}
			Mrk 1044 & wabs*bmc & $0.34_{-0.03}^{+0.03}$ & $0.93_{-0.04}^{+0.05}$ & 0.24 & $1.4_{-0.1}^{+0.1}\times 10^{-4}$ \\
			\noalign{\smallskip}
			RBS 416 & wabs*bmc & $0.26_{-0.27}^{+0.11}$ & $0.95_{-0.08}^{+0.08}$ & 0.73 & $1.4_{-0.3}^{+0.2}\times 10^{-5}$ \\
			\noalign{\smallskip}
			HE 0450-2958 & wabs*(bmc+zgauss) & $0.42_{-0.42}^{+0.10}$ & $1.01_{-0.11}^{+0.15}$ & 0.75 & $2.2_{-0.4}^{+0.1}\times 10^{-5}$ \\
			\noalign{\smallskip}
			PKS 0558-504 & wabs*bmc & $0.34_{-0.03}^{+0.03}$ & $0.96_{-0.03}^{+0.03}$ & 0.34 & $1.9_{-0.1}^{+0.1}\times 10^{-4}$ \\
			\noalign{\smallskip}
			Mrk 110 & wabs*(bmc+zgauss) & $0.30_{-0.02}^{+0.02}$ & $0.64_{-0.03}^{+0.03}$ & 0.50 & $3.0_{-0.1}^{+0.1} \times 10^{-4}$ \\
			\noalign{\smallskip}
			PG 0953+414 & wab*bmc & $0.33_{-0.34}^{+0.10}$ & $0.91_{-0.08}^{+0.11}$ & 0.68 & $3.1_{-0.1}^{+0.1}\times 10^{-5}$ \\
			\noalign{\smallskip}
			RE J1034+396 & wabs*(bmc+zgauss) & $0.31_{-0.11}^{+0.16}$ & $0.90_{-0.50}^{+0.30}$ & 0.04 & $1.3_{-0.2}^{+0.5}\times 10^{-5}$ \\ 
			\noalign{\smallskip}
			PG 1115+407 & wabs*bmc & $0.39_{-0.39}^{+0.09}$ & $0.96_{-0.13}^{+0.09}$ & 0.32 & $1.5_{-0.1}^{+2.5}\times 10^{-5}$ \\
			\noalign{\smallskip}
			PG 1116+215 & wabs*(bmc+zgauss) & $0.31_{-0.08}^{+0.06}$ & $0.82_{-0.04}^{+0.04}$ & 0.85 & $2.9_{-0.1}^{+0.1}\times 10^{-5}$ \\
			\noalign{\smallskip}
			NGC 4051 & wabs*(bmc+zgauss) & $0.44_{-0.07}^{+0.06}$ & $0.54_{-0.05}^{+0.05}$ & 2.00 & $1.0_{-0.1}^{+0.1}\times 10^{-4}$ \\
			\noalign{\smallskip}
			PG 1211+143 & wabs*(bmc+zgauss) & $0.15_{-0.15}^{+0.08}$ & $1.02_{-0.02}^{+0.02}$ & 2.00 & $3.1_{-0.1}^{+0.4}\times 10^{-5}$ \\
			\noalign{\smallskip}
			Mrk 766 & wabs*(bmc+zgauss) & $0.25_{-0.26}^{+0.05}$ & $1.02_{-0.03}^{+0.03}$ & 2.00 & $1.5_{-0.1}^{+0.1}\times 10^{-4}$ \\
			\noalign{\smallskip}
			Was 61 & wabs*bmc & $0.31_{-0.02}^{+0.02}$ & $1.16_{-0.06}^{+0.06}$ & -0.13 & $1.3_{-0.1}^{+0.1}\times 10^{-4}$ \\
			\noalign{\smallskip}
			PG 1244+026 & wabs*bmc & $0.29_{-0.02}^{+0.02}$ & $1.33_{-0.09}^{+0.09}$ & -0.30 & $7.6_{-0.8}^{+0.9}\times 10^{-5}$ \\
			\noalign{\smallskip}
			PG 1322+659 & wabs*bmc & $0.30_{-0.08}^{+0.11}$ & $1.12_{-0.37}^{+0.30}$ & -0.14 & $3.1_{-0.7}^{+1.4}\times 10^{-5}$ \\
			\noalign{\smallskip}
			MCG-6-30-15 & wabs*(bmc+zgauss) & $0.25_{-0.01}^{+0.01}$ & $0.88_{-0.03}^{+0.03}$ & -0.32 & $7.6_{-0.2}^{+0.2} \times 10^{-4}$ \\
			\noalign{\smallskip}
			IRAS 13349+2438 & wabs*(bmc+zgauss) & $0.61_{-0.04}^{+0.04}$ & $1.28_{-0.11}^{+0.13}$ & 2.00 & $3.4_{-0.1}^{+0.1}\times 10^{-5}$ \\
			\noalign{\smallskip}
			PG 1402+261 & wabs*bmc & $0.38_{-0.15}^{+0.11}$ & $0.75_{-0.16}^{+0.22}$ & 0.31 & $1.9_{-0.1}^{+0.2}\times 10^{-5}$ \\
			\noalign{\smallskip}
			NGC 5506 & wabs*(bmc+2zgauss) & $0.38_{-0.02}^{+0.02}$ & $0.76_{-0.01}^{+0.01}$ & 2.00 & $1.1_{-0.1}^{+0.1}\times 10^{-3}$ \\
			\noalign{\smallskip}
			PG 1440+356 & wabs*bmc & $0.30_{-0.10}^{+0.09}$ & $0.96_{-0.13}^{+0.14}$ & 0.30 & $2.4_{-0.2}^{+0.3}\times 10^{-5}$ \\
			\noalign{\smallskip}
			PG 1448+273 & wabs*bmc & $0.28_{-0.03}^{+0.04}$ & $1.30_{-0.19}^{+0.17}$ & -0.45 & $9.8_{-1.9}^{+2.5}\times 10^{-5}$ \\
			\noalign{\smallskip}
			Mrk 493 & wabs*bmc & $0.29_{-0.07}^{+0.05}$ & $0.95_{-0.05}^{+0.05}$ & 0.75 & $1.9_{-0.1}^{+0.1}\times 10^{-5}$ \\
			\noalign{\smallskip}
			IRAS 17020+4544 & wabs*(bmc+zgauss) & $0.26_{-0.03}^{+0.04}$ & $1.15_{-0.05}^{+0.04}$ & 0.34 & $9.4_{-0.6}^{+0.6} \times 10^{-5}$ \\
			\noalign{\smallskip}
			PDS 456 & wabs*(bmc+zgauss) & $0.10_{-0.10}^{+0.28}$ & $1.06_{-0.02}^{+0.02}$ & 1.12 & $3.1_{-0.2}^{+0.7}\times 10^{-5}$ \\
			\noalign{\smallskip}
			Mrk 896 & wabs*bmc & $0.28_{-0.28}^{+0.11}$ & $1.00_{-0.08}^{+0.08}$ & 0.96 & $3.8_{-0.8}^{+1.4}\times 10^{-5}$ \\
			\noalign{\smallskip}
			Mrk 1513 & wabs*(bmc+zgauss) & $0.08_{-0.08}^{+0.16}$ & $0.69_{-0.04}^{+0.04}$ & 0.35 & $2.6_{-1.0}^{+0.9}\times 10^{-5}$ \\
			\noalign{\smallskip}
			II Zw 177 & wabs*bmc & $0.21_{-0.21}^{+0.16}$ & $1.40_{-0.15}^{+0.16}$ & 1.07 & $1.5_{-0.3}^{+7.2}\times 10^{-5}$ \\
			\noalign{\smallskip}
			Ark 564 & wabs*bmc & $0.30_{-0.03}^{+0.02}$ & $1.44_{-0.02}^{+0.03}$ & 0.27 & $4.1_{-0.2}^{-0.3}\times 10^{-4}$ \\
			\noalign{\smallskip}
			AM 2354-304 & wabs*bmc & $0.52_{-0.51}^{+0.18}$ & $1.15_{-0.36}^{+0.60}$ & 0.80 & $4.2_{-1.4}^{+2.6}\times 10^{-5}$ \\
			\bottomrule
		\end{tabular}
	\end{center}
	\begin{flushleft}
		Columns: 1 = AGN name. 
		2 = \textsc{xspec} model used. 3 = temperature of thermal seed photons in keV. 4 = energy spectral index ($\alpha = \Gamma - 1$). 5 = logarithm of the $A$ parameter (which is related to the Comptonization fraction $f$ by $f = A/(1+A)$). 6 = BMC normalization. Numbers in columns 3--6 without error ranges were frozen at their best-fit values.
	\end{flushleft}
	\label{tab:spectralnls1}
\end{table*}  

\begin{table*}
	\caption{BLS1 spectral data}
	\begin{center}
		\begin{tabular}{lccccc} 
			\toprule
			\toprule       
			Name & Model &  $kT$ &  $\alpha$ &  log $A$ &  Norm \\
			(1) & (2) & (3) & (4) & (5) & (6) \\
			\midrule
			PG 0052+251 & wabs*bmc & $0.28_{-0.16}^{+0.11}$ & $0.83_{-0.06}^{+0.15}$ & 0.53 & $5.7_{-2.5}^{+0.8}\times 10^{-5}$ \\
			\noalign{\smallskip}
			Q 0056-363 & wabs*bmc & $0.35_{-0.09}^{+0.08}$ & $0.78_{-0.09}^{+0.11}$ & 0.36 & $3.1_{-0.1}^{+0.1}\times 10^{-5}$ \\
			\noalign{\smallskip}
			Mrk 1152 & wabs*(bmc+zgauss) & $0.33_{-0.09}^{+0.10}$ & $0.49_{-0.10}^{+0.08}$ & 0.61 & $4.4_{-0.6}^{+0.8}\times 10^{-5}$ \\
			\noalign{\smallskip}
			ESO 244-G17 & wabs*bmc & $0.24_{-0.24}^{+0.06}$ & $0.68_{-0.10}^{+0.09}$ & 0.40 & $3.3_{-1.9}^{+0.1}\times 10^{-5}$ \\
            \noalign{\smallskip}
			Fairall 9 & wabs*(bmc+zgauss) & $0.48_{-0.48}^{+0.10}$ & $0.81_{-0.14}^{+0.17}$ & 0.59 & $2.4_{-1.0}^{+0.1}\times 10^{-4}$ \\
			\noalign{\smallskip}
			Mrk 590 & wabs*(bmc+zgauss) & $0.24_{-0.06}^{+0.04}$ & $0.57_{-0.06}^{+0.06}$ & 0.36 & $6.4_{-0.5}^{+0.3}\times 10^{-5}$ \\
			\noalign{\smallskip}
			ESO 198-G24 & wabs*(bmc+zgauss) & $0.31_{-0.03}^{+0.03}$ & $0.59_{-0.03}^{+0.03}$ & 0.52 & $1.0_{-0.1}^{+0.1}\times 10^{-4}$ \\
			\noalign{\smallskip}
			Fairall 1116 & wabs*(bmc+zgauss) & $0.26_{-0.11}^{+0.09}$ & $0.81_{-0.07}^{+0.06}$ & 0.58 & $5.3_{-0.3}^{+0.2}\times 10^{-5}$ \\
			\noalign{\smallskip}
			1H 0419-577 & wabs*bmc & $0.30_{-0.02}^{+0.02}$ & $0.59_{-0.04}^{+0.03}$ & 0.40 & $1.4_{-0.1}^{+0.1}\times 10^{-4}$ \\
			\noalign{\smallskip}
			3C 120 & wabs*(bmc+2zgauss) & $0.33_{-0.02}^{+0.03}$ & $0.67_{-0.03}^{+0.03}$ & 0.46 & $4.9_{-0.1}^{+0.1}\times 10^{-4}$ \\
			\noalign{\smallskip}
			H 0439-272 & wabs*bmc & $0.40_{-0.13}^{+0.07}$ & $0.70_{-0.06}^{+0.10}$ & 0.52 & $6.1_{-0.6}^{+0.8}\times 10^{-5}$ \\
			\noalign{\smallskip}
			MCG-01-13-25 & wabs*bmc & $0.29_{-0.09}^{+0.09}$ & $0.55_{-0.16}^{+0.11}$ & 0.39 & $1.3_{-0.1}^{+0.2}\times 10^{-4}$ \\
			\noalign{\smallskip}
			Ark 120 & wabs*(bmc+2zgauss) & $0.28_{-0.01}^{+0.01}$ & $0.78_{-0.02}^{+0.02}$ & 0.20 & $4.6_{-0.1}^{+0.1}\times 10^{-4}$ \\
			\noalign{\smallskip}
			MCG-02-14-09 & wabs*(bmc+zgauss) & $0.36_{-0.04}^{+0.03}$ & $0.73_{-0.03}^{+0.04}$ & 0.54 & $4.4_{-0.1}^{+0.1}\times 10^{-5}$ \\
			\noalign{\smallskip}
			MCG+8-11-11 & wabs*(bmc+zgauss) & $0.36_{-0.05}^{+0.05}$ & $0.54_{-0.03}^{+0.04}$ & 0.66 & $4.4_{-0.3}^{+0.3}\times 10^{-4}$ \\
			\noalign{\smallskip}
			H 0557-385 & wabs*bmc & $0.09_{-0.09}^{+0.37}$ & $0.71_{-0.06}^{+0.06}$ & 0.43 & $3.4_{-1.1}^{+2.0}\times 10^{-4}$ \\
            \noalign{\smallskip}
			PMN J0623-6436 & wabs*bmc & $0.43_{-0.43}^{+0.12}$ & $0.62_{-0.16}^{+0.25}$ & 0.50 & $3.9_{-2.2}^{+0.6}\times 10^{-5}$ \\
			\noalign{\smallskip}
			ESO 209-G12 & wabs*(bmc+zgauss) & $0.22_{-0.22}^{+0.10}$ & $0.73_{-0.10}^{+0.07}$ & 0.47 & $7.9_{-4.2}^{+0.5}\times 10^{-5}$ \\
			\noalign{\smallskip}
			PG 0804+761 & wabs*(bmc+zgauss) & $0.28_{-0.14}^{+0.06}$ & $1.01_{-0.04}^{+0.04}$ & 0.82 & $9.6_{-0.2}^{+0.4}\times 10^{-5}$ \\
			\noalign{\smallskip}
			Fairall 1146 & wabs*(bmc+zgauss) & $0.25_{-0.24}^{+0.08}$ & $0.80_{-0.09}^{+0.08}$ & 0.55 & $1.3_{-0.8}^{+0.1}\times 10^{-4}$ \\
			\noalign{\smallskip}
			PG 0844+349 & wabs*(bmc+zgauss) & $0.22_{-0.22}^{+0.42}$ & $0.23_{-0.23}^{+0.35}$ & 0.44 & $9.1_{-7.1}^{+17.0}\times 10^{-6}$ \\
			\noalign{\smallskip}
			MCG+04-22-42 & wabs*(bmc+zgauss) & $0.41_{-0.08}^{+0.06}$ & $0.67_{-0.06}^{+0.07}$ & 0.51 & $1.7_{-0.1}^{+0.1}\times 10^{-4}$ \\
			\noalign{\smallskip}
			PG 0947+396 & wabs*(bmc+zgauss) & $0.31_{-0.13}^{+0.14}$ & $0.64_{-0.19}^{+0.14}$ & 0.44 & $1.8_{-0.1}^{+0.3}\times 10^{-5}$ \\
			\noalign{\smallskip}
			HE 1029-1401 & wabs*(bmc+zgauss) & $0.28_{-0.02}^{+0.02}$ & $0.83_{-0.04}^{+0.03}$ & 0.96 & $1.7_{-0.1}^{+0.1}\times 10^{-4}$ \\
			\noalign{\smallskip}
			PG 1048+342 & wabs*(bmc+zgauss) & $0.32_{-0.32}^{+0.13}$ & $0.70_{-0.09}^{+0.10}$ & 0.76 & $1.3_{-1.3}^{+0.1}\times 10^{-5}$ \\
			\noalign{\smallskip}
			NGC 3516 & wabs*(bmc+zgauss) & $0.47_{-0.02}^{+0.02}$ & $0.73_{-0.02}^{+0.02}$ & 2.00 & $4.1_{-0.1}^{+0.1}\times 10^{-4}$ \\
			\noalign{\smallskip}
			PG 1114+445 & wabs*bmc & $0.11_{-0.11}^{+0.12}$ & $0.38_{-0.08}^{+0.08}$ & 0.49 & $1.2_{-1.2}^{+0.8}\times 10^{-5}$ \\
			\noalign{\smallskip}
			NGC 3783 & wabs*(bmc+zgauss) & $0.37_{-0.02}^{+0.02}$ & $0.50_{-0.01}^{+0.01}$ & 2.00 & $4.8_{-0.2}^{+0.1}\times 10^{-4}$ \\
			\noalign{\smallskip}
			HE 1143-1810 & wabs*(bmc+zgauss) & $0.34_{-0.03}^{+0.03}$ & $0.65_{-0.04}^{+0.04}$ & 0.36 & $3.2_{-0.1}^{+0.1}\times 10^{-4}$ \\
			\noalign{\smallskip}
			PG 1202+281 & wabs*bmc & $0.38_{-0.13}^{+0.11}$ & $0.59_{-0.15}^{+0.15}$ & 0.45 & $3.9_{-0.3}^{+0.5}\times 10^{-5}$ \\
            \noalign{\smallskip}
			Mrk 205 & wabs*(bmc+zgauss) & $0.14_{-0.15}^{+0.07}$ & $1.00_{-0.02}^{+0.02}$ & 2.00 & $1.1_{-0.1}^{+0.1}\times 10^{-4}$ \\
			\noalign{\smallskip}
			Ark 374 & wabs*bmc & $0.36_{-0.07}^{+0.07}$ & $0.79_{-0.16}^{+0.16}$ & 0.09 & $4.9_{-0.2}^{+0.4}\times 10^{-5}$ \\
			\noalign{\smallskip}
			NGC 4593 & wabs*(bmc+zgauss) & $0.38_{-0.05}^{+0.04}$ & $0.57_{-0.05}^{+0.05}$ & 0.48 & $2.7_{-0.1}^{+0.1}\times 10^{-4}$ \\
			\noalign{\smallskip}
			PG 1307+085 & wabs*(bmc+zgauss) & $0.39_{-0.39}^{+0.19}$ & $0.59_{-0.13}^{+0.22}$ & 2.00 & $1.5_{-1.5}^{+0.3}\times 10^{-5}$ \\
            \noalign{\smallskip}
			4U 1344-60 & wabs*bmc & $0.36_{-0.36}^{+0.12}$ & $0.51_{-0.06}^{+0.07}$ & 0.85 & $5.4_{-5.4}^{+0.9}\times 10^{-5}$ \\
			\noalign{\smallskip}
			IC 4329A & wabs*(bmc+2zgauss) & $0.36_{-0.02}^{+0.02}$ & $0.63_{-0.01}^{+0.01}$ & 0.59 & $9.9_{-0.1}^{+0.1}\times 10^{-4}$ \\
			\noalign{\smallskip}
			Mrk 279 & wabs*(bmc+zgauss) & $0.38_{-0.06}^{+0.05}$ & $0.64_{-0.05}^{+0.06}$ & 0.47 & $2.6_{-0.1}^{+0.1}\times 10^{-4}$ \\
			\noalign{\smallskip}
			PG 1352+183 & wabs*bmc & $0.26_{-0.02}^{+0.02}$ & $1.14_{-0.27}^{+0.26}$ & -1.17 & $3.7_{-0.9}^{+1.4}\times 10^{-4}$ \\
            \noalign{\smallskip}
			PG 1415+451 & wabs*(bmc+zgauss) & $0.37_{-0.18}^{+0.11}$ & $0.87_{-0.18}^{+0.27}$ & 0.23 & $1.4_{-0.1}^{+0.4}\times 10^{-5}$ \\
            \noalign{\smallskip}
			NGC 5548 & wabs*(bmc+2zgauss) & $0.24_{-0.01}^{+0.01}$ & $0.45_{-0.02}^{+0.02}$ & 0.23 & $3.0_{-0.1}^{+0.1}\times 10^{-4}$ \\
			\noalign{\smallskip}
			PG 1416-129 & wabs*bmc & $0.36_{-0.03}^{+0.03}$ & $0.54_{-0.14}^{+0.13}$ & -0.02 & $5.5_{-0.1}^{+0.3}\times 10^{-5}$ \\
			\noalign{\smallskip}
			PG 1425+267 & wabs*bmc & $0.47_{-0.47}^{+0.13}$ & $0.57_{-0.13}^{+0.16}$ & 2.00 & $1.5_{-1.0}^{+0.2}\times 10^{-5}$ \\
			\noalign{\smallskip}
			Mrk 1383 & wabs*(bmc+zgauss) & $0.50_{-0.50}^{+0.10}$ & $0.79_{-0.16}^{+0.19}$ & $0.49$ & $9.6_{-0.5}^{+0.6}\times 10^{-5}$ \\
			\bottomrule
		\end{tabular}
	\end{center}
	\begin{flushleft}
		Columns: 1 = AGN name. 
		2 = \textsc{xspec} model used. 3 = temperature of thermal seed photons in keV. 4 = energy spectral index ($\alpha = \Gamma - 1$). 5 = logarithm of the $A$ parameter (which is related to the Comptonization fraction $f$ by $f = A/(1+A)$). 6 = BMC normalization. Numbers in columns 3--6 without error ranges were frozen at their best-fit values.
	\end{flushleft}
	\label{tab:spectralbls1_part1}
\end{table*}  

\setcounter{table}{1}
\begin{table*}
	\caption{BLS1 spectral data (\textit{continued})}
		\centering
	\begin{center}
		\begin{tabular}{lccccc} 
			\toprule
			\toprule       
			Name & Model &  $kT$ &  $\alpha$ &  log $A$ &  Norm \\
			(1) & (2) & (3) & (4) & (5) & (6) \\
			\midrule
			PG 1427+480 & wabs*bmc & $0.33_{-0.05}^{+0.05}$ & $0.93_{-0.22}^{+0.20}$ & -0.22 & $2.8_{-0.3}^{+0.5}\times 10^{-5}$ \\
            \noalign{\smallskip}
			Mrk 841 & wabs*bmc & $0.25_{-0.05}^{+0.05}$ & $0.75_{-0.08}^{+0.07}$ & 0.31 & $1.8_{-0.1}^{+0.1}\times 10^{-4}$ \\
            \noalign{\smallskip}
			Mrk 290 & wabs*(bmc+zgauss) & $0.23_{-0.25}^{+0.11}$ & $0.56_{-0.10}^{+0.10}$ & 0.47 & $6.8_{-4.7}^{+1.3}\times 10^{-5}$ \\
			\noalign{\smallskip}
			Mrk 876 & wabs*bmc & $0.30_{-0.03}^{+0.02}$ & $0.50_{-0.33}^{+0.44}$ & -0.99 & $3.7_{-0.3}^{+1.0}\times 10^{-4}$ \\
			\noalign{\smallskip}
			PG 1626+554 & wabs*bmc & $0.33_{-0.13}^{+0.09}$ & $0.50_{-0.35}^{+0.38}$ & 0.14 & $3.9_{-0.3}^{+2.4} \times 10^{-5}$ \\
			\noalign{\smallskip}
			IGR J17418-1212 & wabs*bmc & $0.40_{-0.40}^{+0.11}$ & $0.77_{-0.16}^{+0.12}$ & 0.49 & $1.5_{-0.1}^{+0.1} \times 10^{-4}$ \\
			\noalign{\smallskip}
			Mrk 509 & wabs*bmc & $0.36_{-0.02}^{+0.02}$ & $0.61_{-0.02}^{+0.02}$ & 0.39 & $5.7_{-0.1}^{+0.1} \times 10^{-4}$ \\
			\noalign{\smallskip}
			Mrk 304 & wabs*(bmc+zgauss) & $0.05_{-0.05}^{+1.19}$ & $0.79_{-0.18}^{+0.19}$ & 1.26 & $2.4_{-1.0}^{+0.1} \times 10^{-5}$ \\
			\noalign{\smallskip}
			MR 2251-178 & wabs*bmc & $0.33_{-0.05}^{+0.06}$ & $0.53_{-0.04}^{+0.03}$ & 0.70 & $4.2_{-0.2}^{+0.3} \times 10^{-4}$ \\
			\noalign{\smallskip}
			NGC 7469 & wabs*(bmc+zgauss) & $0.39_{-0.02}^{+0.02}$ & $0.62_{-0.02}^{+0.02}$ & 0.47 & $3.4_{-0.1}^{+0.1} \times 10^{-4}$ \\
			\noalign{\smallskip}
			Mrk 926 & wabs*bmc & $0.24_{-0.05}^{+0.04}$ & $0.60_{-0.05}^{+0.04}$ & 0.38 & $3.0_{-0.2}^{+0.1} \times 10^{-4}$ \\
			\bottomrule
		\end{tabular}
	\end{center}
	\begin{flushleft}
		Columns: 1 = AGN name. 2 = \textsc{xspec} model used. 3 = temperature of thermal seed photons in keV. 4 = energy spectral index ($\alpha = \Gamma - 1$). 5 = logarithm of the $A$ parameter (which is related to the Comptonization fraction $f$ by $f = A/(1+A)$). 6 = BMC normalization. Numbers in columns 3--6 without error ranges were frozen at their best-fit values.
	\end{flushleft}
	\label{tab:spectralbls1_part2}
\end{table*}

Based on our spectral analysis, NLS1s have 2--10 keV luminosity values that range from $1.4 \times 10^{41}$ erg s$^{-1}$ to $7.6 \times 10^{44}$ erg s$^{-1}$ with a mean of $9.7 \times 10^{43}$ erg s$^{-1}$. BLS1s, on the other hand, have \lx\ values that range from $2.1 \times 10^{42}$ erg s$^{-1}$ to $6.0 \times 10^{44}$ erg s$^{-1}$ with a mean of $1.1 \times 10^{44}$ erg s$^{-1}$.

In summary, our homogeneous analysis of NLS1s and BLS1s indicates that the two samples have the same characteristics based on X-ray luminosity and spectral parameters, with the notable exception of the spectral index, which is significantly steeper in NLS1s.

\section{BH Mass and accretion rate estimates}
\begin{figure}
	\includegraphics[width=\columnwidth]{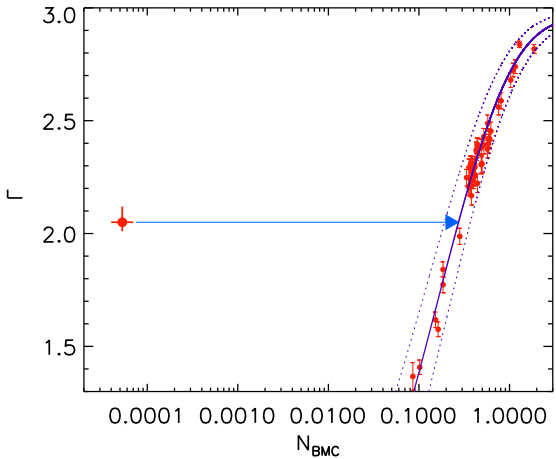}
	\caption{\nbmc--$\Gamma$ diagram illustrating the scaling method for an AGN (in this case, Ton S180), represented by the large filled-in circle on the left-hand side, with the reference source \xte, whose spectral trend is fitted by the solid curve on the right-hand side. The dotted lines indicate the 1-sigma uncertainties on the best-fit function.}
	\label{fig:scaling}
\end{figure}

We determined the black hole mass of the AGNs in our sample by applying the X-ray scaling method, which is described in detail in Gliozzi et al. (2011). The method's basic assumption is that the physics of black hole systems is scale-invariant, which means we can calculate the BH mass of an AGN by comparing it to a known stellar-mass, Galactic black hole (GBH) that we use as a reference source. More specifically, the method assumes that black hole systems undergo the same X-ray spectral evolution associated with changes in accretion rates. For stellar-mass black holes the evolution from low-hard state to high-soft state and vice versa is well documented and can be illustrated by the similar trends shown by different sources when $\Gamma$ is plotted versus \nbmc\ \citep{shap09}. For AGNs, the timescales for spectral evolution are much longer; therefore, there is only one data point in the \nbmc-$\Gamma$ plot, and it can be directly compared to the trend shown by stellar-mass black holes. This procedure is illustrated in Figure~\ref{fig:scaling}, where the large filled-in circle on the left side represents the location of the AGN (in this example, the NLS1 Ton S180) in the \nbmc--$\Gamma$ plot, and the data points on the right side represent the spectral evolution of the reference source \xte\ during the 1998 outburst, which are well fitted by a function indicated by the continuous line (the mathematical expression of this best-fit function along with the fitting functions of other reference sources are reported in Gliozzi et al. 2011). The scaling factor is provided by the ratio \nbmct/\nbmcr, which is illustrated by the horizontal arrow in Figure~\ref{fig:scaling}. The \textit{t} subscript denotes the target AGN, and the \textit{r} subscript denotes the reference source. Since \nbmc\ is directly proportional to the accretion luminosity of the BH system, which in turns depends on \mbh\ and \mdot, and since $\Gamma$ defines the accretion state of the source, comparing the \nbmc\ values of target and reference, obtained for the same value of $\Gamma$, yields the scaling factor for the BH mass.
In this method the statistical errors on \mbh\ depend on the uncertainties of the AGN position (illustrated by the vertical and horizontal error bars in Figure~\ref{fig:scaling}) as well as on the uncertainties of the functional representation (illustrated by the dotted lines in the figure). For the values of log(\mbh) of the AGNs in this work, derived using \xte\ as reference source (reported in Tables~\ref{tab:nls1s} and \ref{tab:bls1s}), the average error is 0.3 dex.

The black hole mass of each AGN is determined with the following equation:

\begin{equation}
M_{\mathrm{BH,t}}=M_{\mathrm{BH,r}} \times \frac {N_{\mathrm{BMC,t}}} {N_{\mathrm{BMC,r}}} \times \left(\frac{d_{\mathrm{t}}}{d_{\mathrm{r}}}\right)^2
\end{equation}
\\
where $d_\mathrm{t}/d_\mathrm{r}$ is the ratio of the distances to the target AGN and the reference source.

Tables \ref{tab:nls1s} and \ref{tab:bls1s} list the black hole masses obtained with this X-ray scaling method. The RM column in the tables is the \mbh\ obtained by reverberation mapping and is included for comparison.

When applying the X-ray scaling method to AGNs, we cannot know a priori which of the reference spectral trends better represent the spectral evolution of a specific AGN, since the latter yields only one isolated point in the \nbmc--$\Gamma$ plot. However, only the spectral trend of the reference \xte\ can be compared to  NLS1s that are characterized by steep $\Gamma$ values. Therefore, we regard \xte\ as our primary reference source in this work, since it allows the determination of \mbh\ for all the objects in our samples. For completeness we also use two other reference sources, \gro\ and \gx, and report the \mbh\ values obtained with those reference sources as well. The similarity of the \mbh\ values obtained with different reference sources demonstrates that our conclusions do not depend on the specific choice of reference source. Note that based on the analysis from Gliozzi et al. (2011), all reference sources are consistent with each other, but \xte\ has a tendency to provide \mbh\ systematically lower by a factor of three, compared to the values obtained with \gro\ and \gx. Therefore, the \mbh\ values reported in Tables \ref{tab:nls1s} and \ref{tab:bls1s} obtained using the reference source \xte\ have been increased by a factor of three.

This method was systematically applied to all AGNs in our samples. The only exceptions are the AGNs that have $\Gamma$ with unphysically low values, which cannot be compared with any of our reference sources. These are AGNs whose X-ray emission is dominated by reflection or absorption or both and hence cannot be compared with the primary coronal emission of the reference sources. As we mentioned in Section 3, nine of the AGNs were in this category, leaving us with a total of 89 AGNs.

Next we move to a comparison of NLS1s and BLS1s by accretion rate rather than \mbh, and our results are shown in Tables \ref{tab:nls1s_ar} and \ref{tab:bls1s_ar}. We used two proxies for accretion rate. The first is the ratio of the X-ray luminosity \lx\ derived in our spectral analysis to the Eddington luminosity \ledd\ obtained from the \mbh\ values. The second is the ratio of bolometric luminosity \lbol\ to \ledd. To determine the bolometric luminosity of all our objects, we used the correction factors derived by Vasudevan and Fabian (2009) by fitting broadband spectral energy distributions (SEDs) obtained with simultaneous data from a sizable sample of AGNs. More specifically, for objects in our samples that are also in the Vasudevan and Fabian sample, we used the specific bolometric correction values derived from their SED fitting. For all the other objects, we used average correction factors that we obtained by averaging separately the NLS1s and BLS1s in the Vasudevan and Fabian sample. For NLS1s, the average bolometric correction is 96.4, whereas for BLS1s it is 23.7.

Tables \ref{tab:nls1s}--\ref{tab:bls1s_ar}, then, list the three numbers, \mbh, \lx/\ledd, and \lbol/\ledd, for each AGN using all three reference sources. An ellipsis in the tables denotes that the value of $\Gamma$ for that AGN, as derived by the \textsc{xspec} fit, fell outside the known range of $\Gamma$ for the reference source and, therefore, that source could not be used in the scaling process to determine \mbh\ for that particular AGN.

The results of our analysis are illustrated in Figures~\ref{fig:histmbh}--\ref{fig:histlbollEdd}. Figure~\ref{fig:histmbh} shows the distribution of NLS1s and BLS1s by black hole mass. The number of NLS1s is shown by a solid line (blue in the online version), and the number of BLS1s, by a dashed line (red online). Figure~\ref{fig:histlxlEdd} shows the distribution of NLS1s and BLS1s by the ratio \lx/\ledd. Finally, Figure~\ref{fig:histlbollEdd} compares NLS1s and BLS1s by \lbol/\ledd. A visual inspection of Figures~\ref{fig:histmbh}--\ref{fig:histlbollEdd} suggests that the distributions of \mbh, \lx/\ledd, and \lbol/\ledd\ of NLS1s are different from the respective distributions of BLS1s. This conclusion was confirmed by statistical tests, whose results are summarized in Table~\ref{tab:stats}. The top half of the table shows the differences between NLS1s and BLS1s in their average values for each parameter. Each number in the bottom half of the table is the probability of finding a difference this large given the null hypothesis that our samples of NLS1s and BLS1s were taken from the same population. On average the \mbh\ values of NLS1s are smaller than the corresponding values of BLS1s. The difference between the \mbh\ averages is marginally significant at the 2.6 sigma level, and a t-test indicates that the average \mbh\ values of NLS1s and BLS1s are consistent with being drawn from the same population with a probability of 5\%. A difference at higher significance level between the two AGN populations is obtained using a nonparametric Kolmogorov--Smirnov (K--S) test, which indicates that the probability that the distributions of \mbh\ from NLS1s and BLS1s are drawn from the same population is 1\%. Differences between NLS1s and BLS1s at even higher significance levels are inferred when the accretion rates are compared: the distributions of both accretion rate indicators---\lx/\ledd\ and \lbol/\ledd---have less than a 0.01\% probability of being drawn from the same population.

\section{Discussion and conclusions}

In this work, we have carried out a homogeneous spectral analysis of the X-ray data with an absorbed Comptonization model (BMC) of two samples of NLS1s and BLS1s, derived from a flux-limited sample of type I AGNs observed with the \xmm\ satellite. Starting from the spectral-fitting results, we applied an X-ray scaling method to infer the BH masses and accretion rate values.

\begin{figure*}
    \includegraphics[width=0.69\columnwidth]{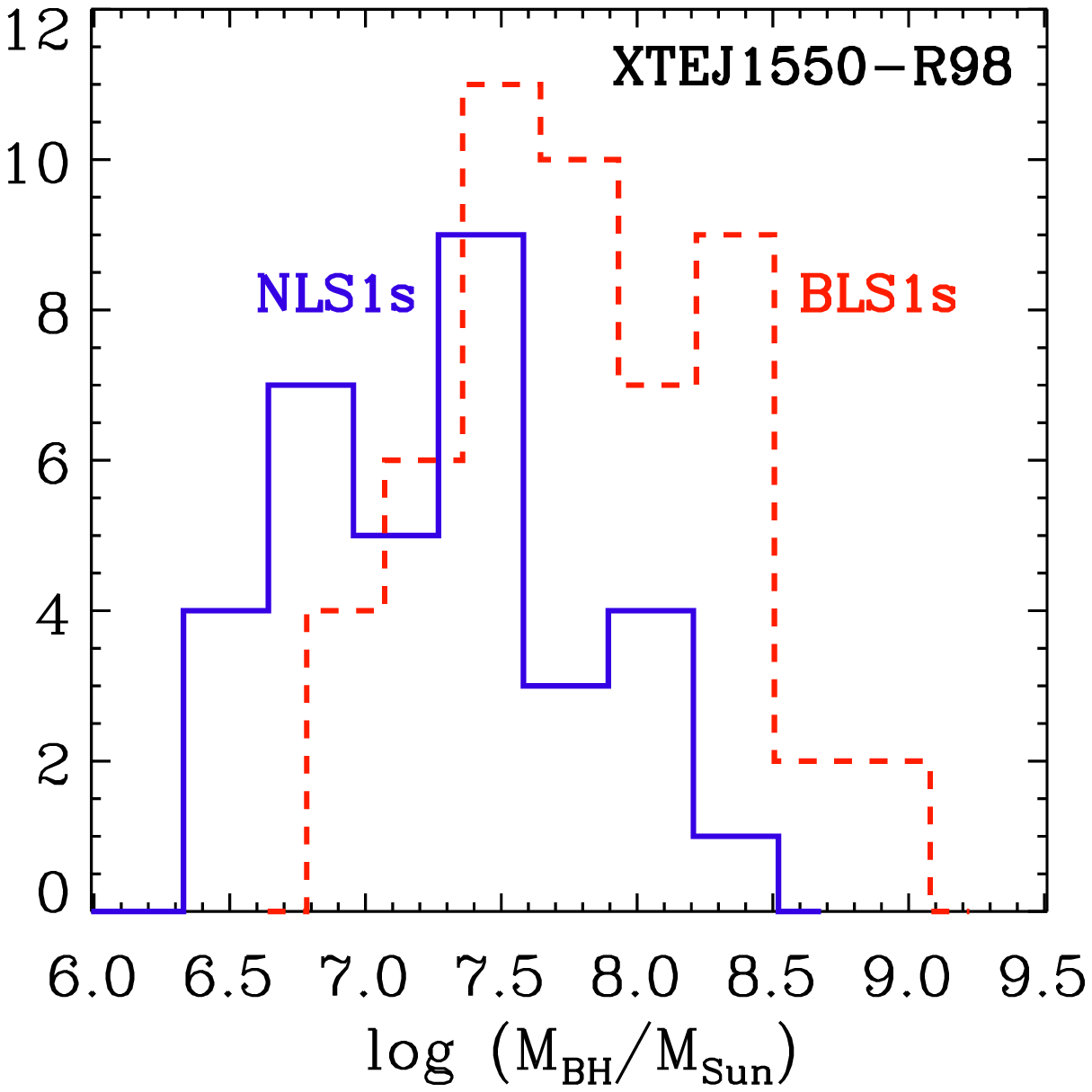}
    \hfill
	\includegraphics[width=0.69\columnwidth]{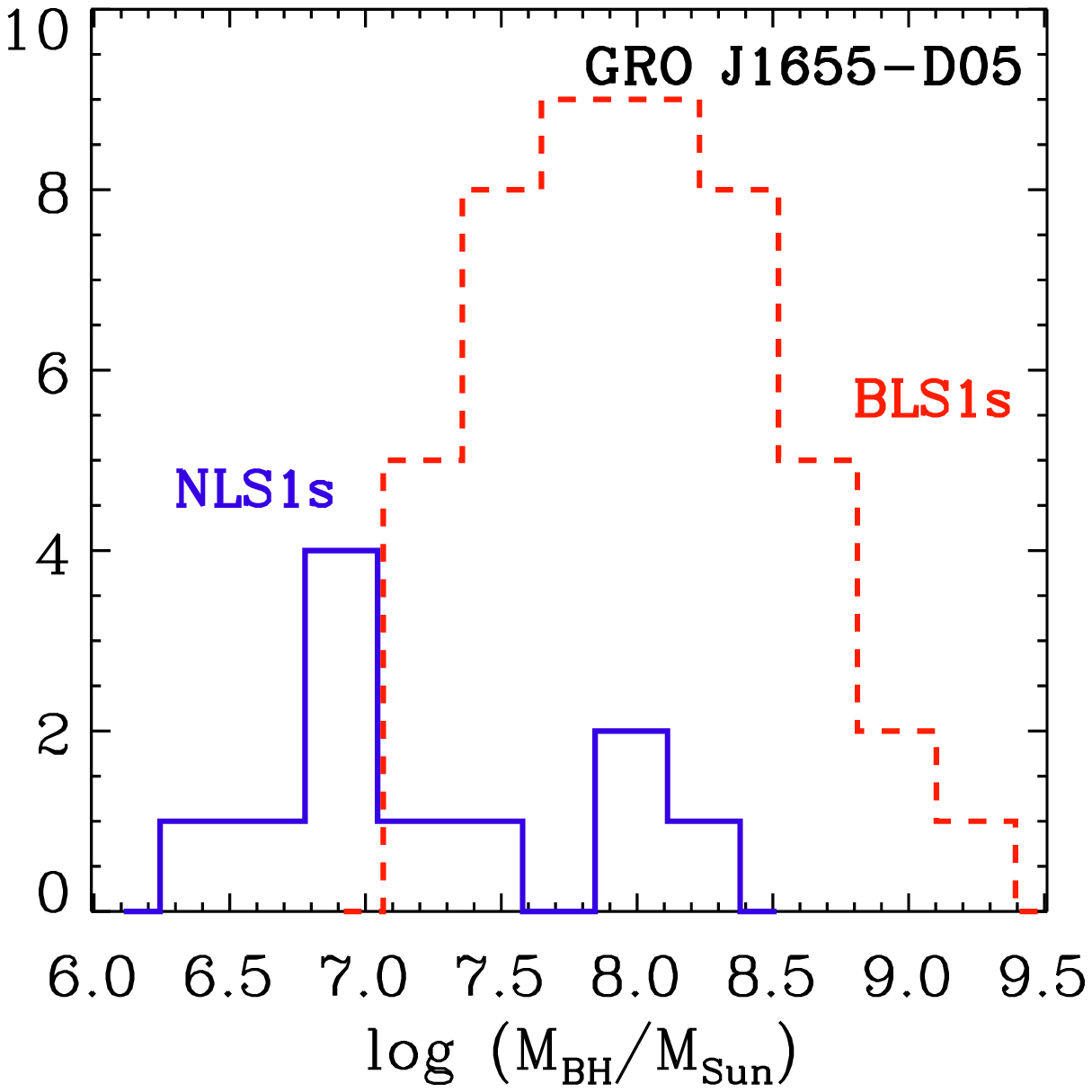}
    \hfill
	\includegraphics[width=0.69\columnwidth]{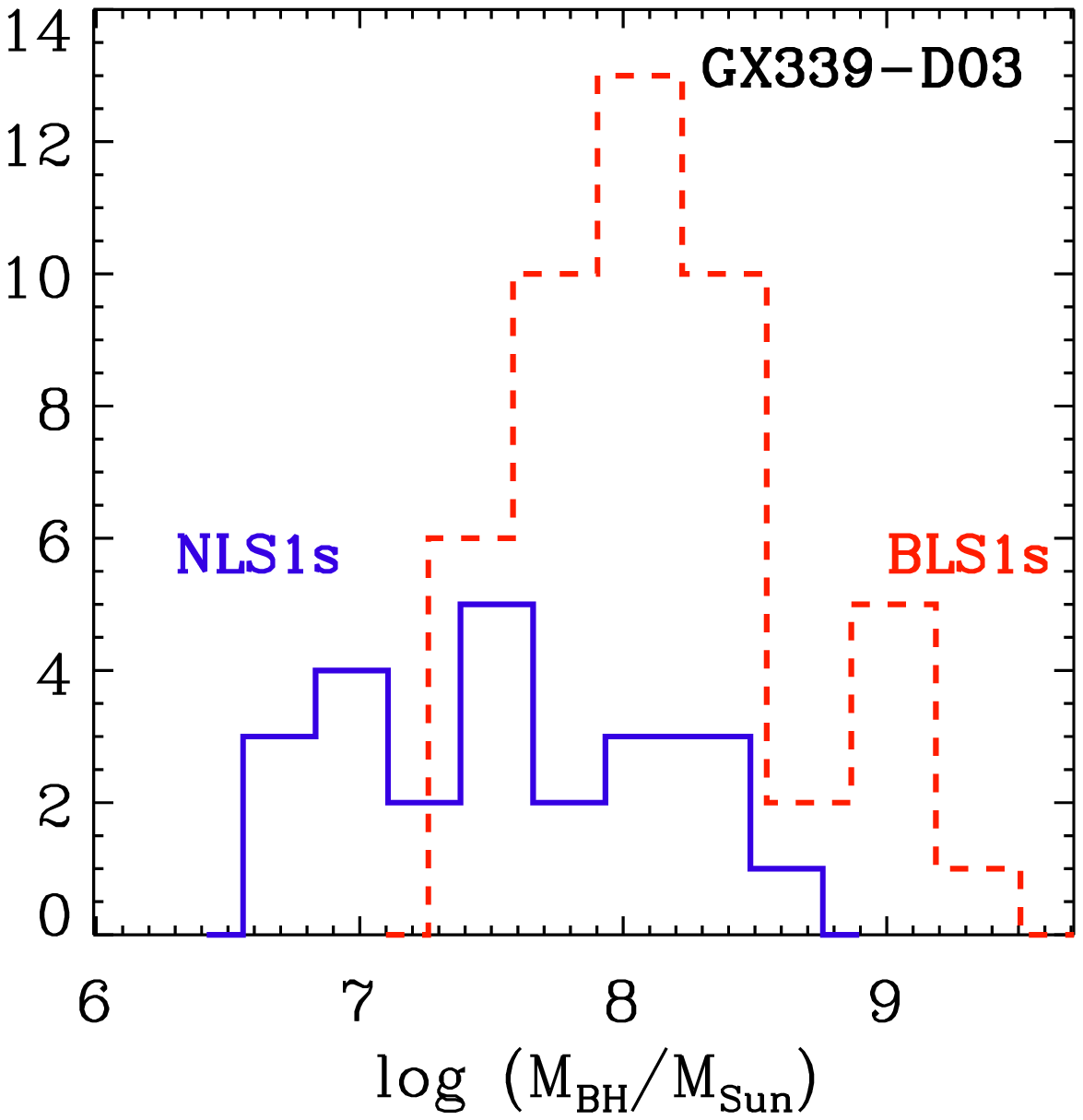}
    \caption{Number of NLS1s and BLS1s by black hole mass using reference source \xte\ (\textit{left}), \gro\ (\textit{center}), and \gx\ (\textit{right}).}
    \label{fig:histmbh}
\end{figure*}

\begin{figure*}
   	\includegraphics[width=0.69\columnwidth]{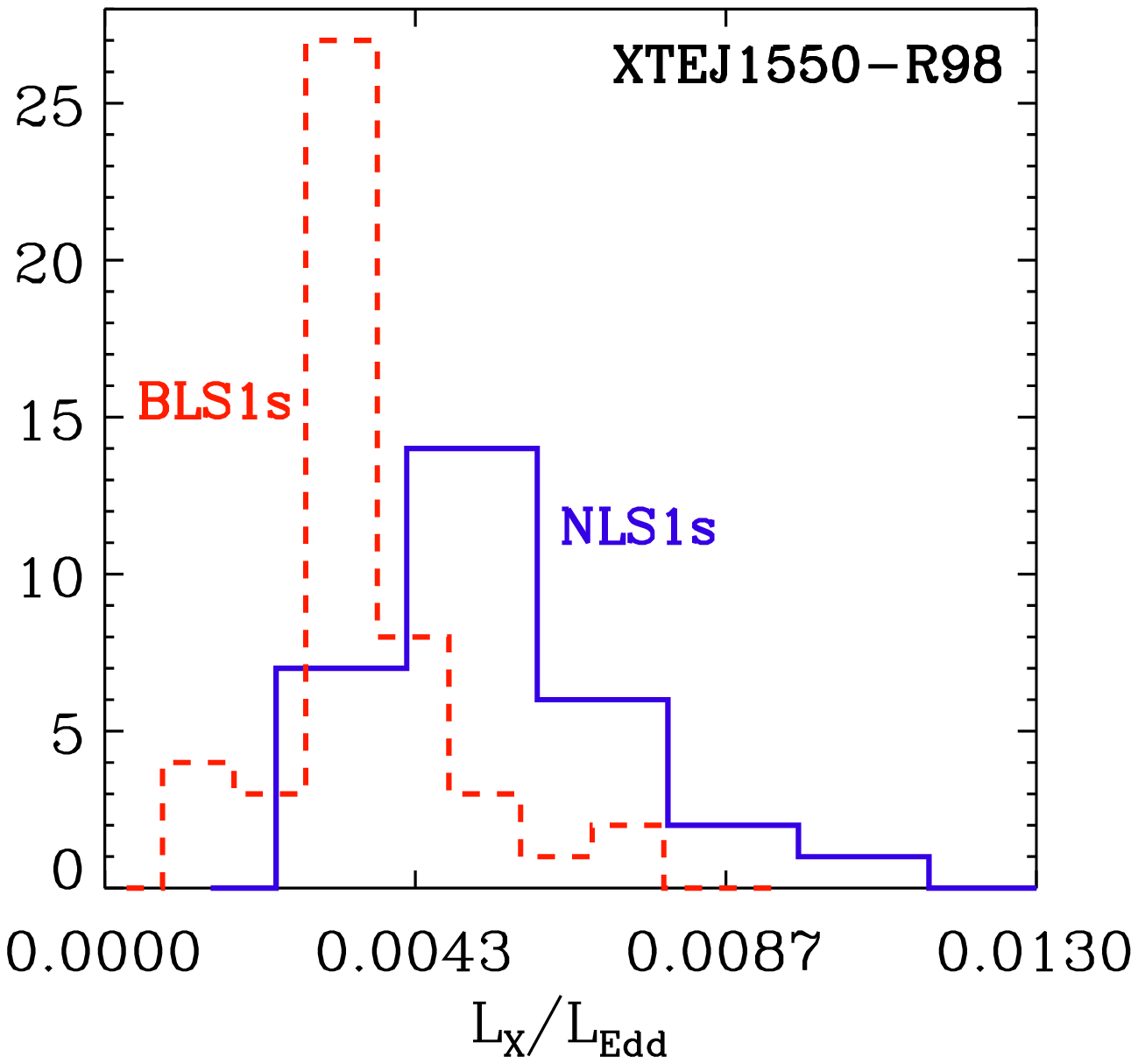}
    \hfill
   	\includegraphics[width=0.69\columnwidth]{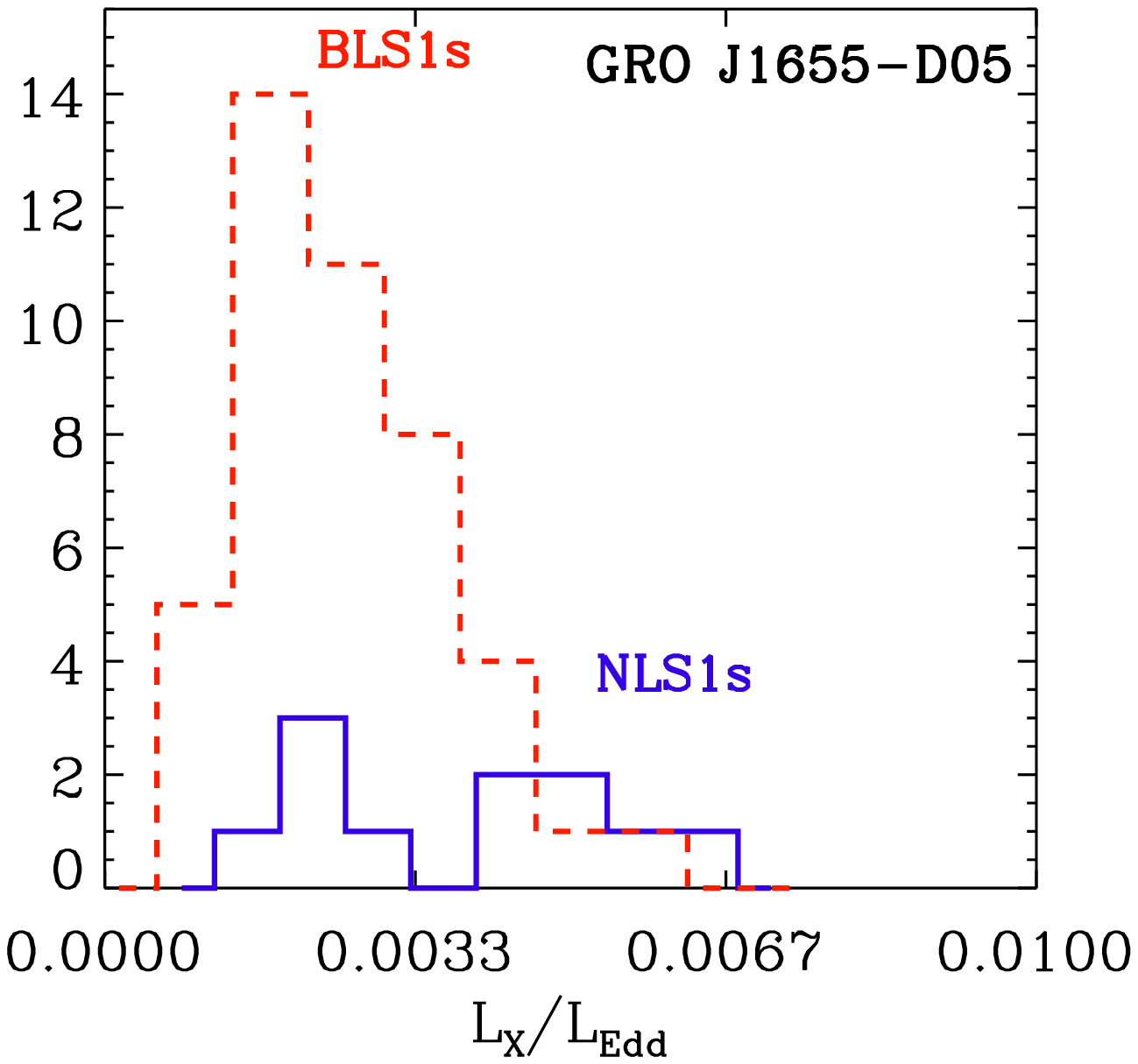}
    \hfill
   	\includegraphics[width=0.69\columnwidth]{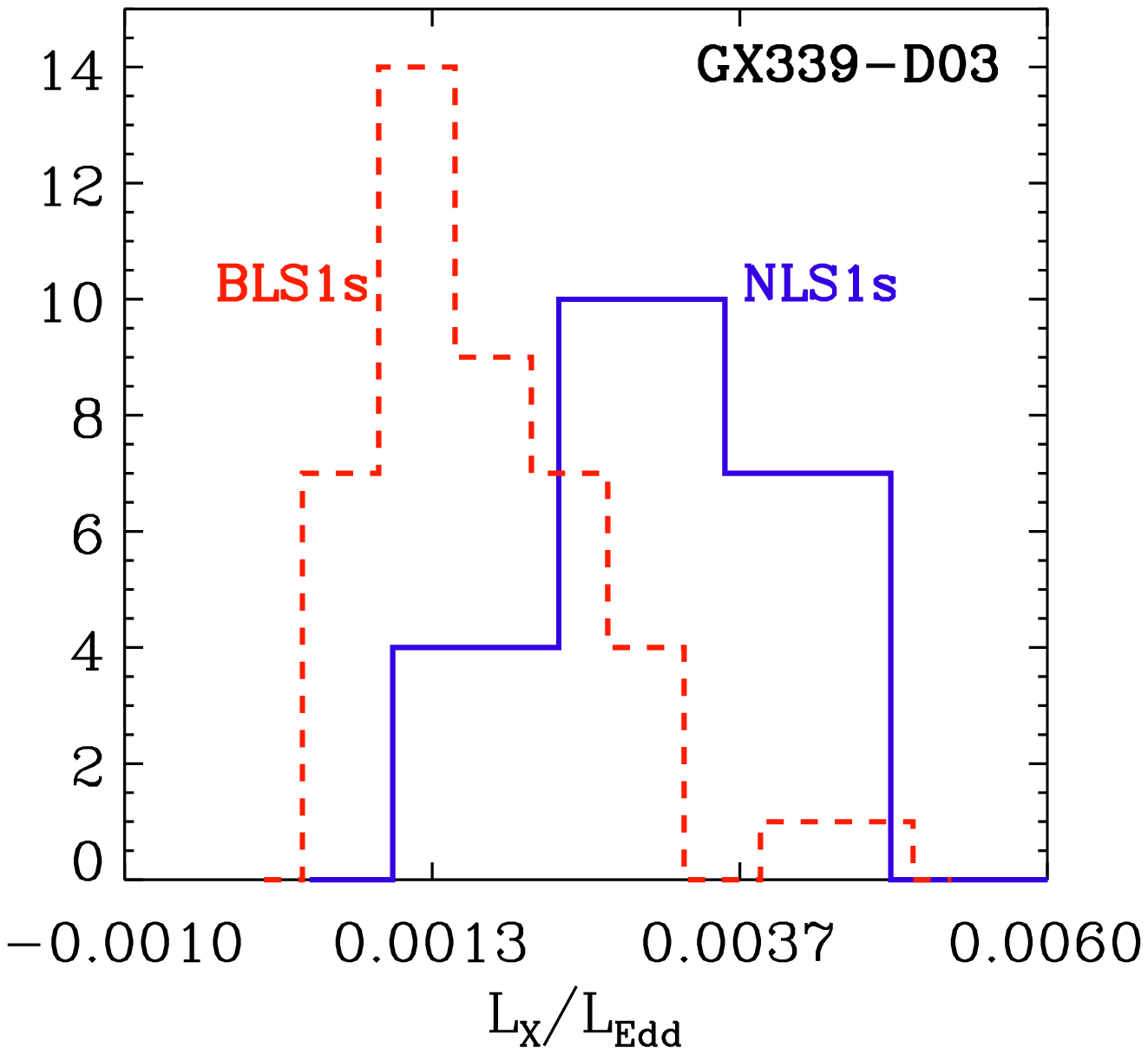}
    \caption{Number of NLS1s and BLS1s by \lx/\ledd\ using reference source \xte\ (\textit{left}), \gro\ (\textit{center}), and \gx\ (\textit{right}).}
    \label{fig:histlxlEdd}
\end{figure*}

\begin{figure*}
	\includegraphics[width=0.69\columnwidth]{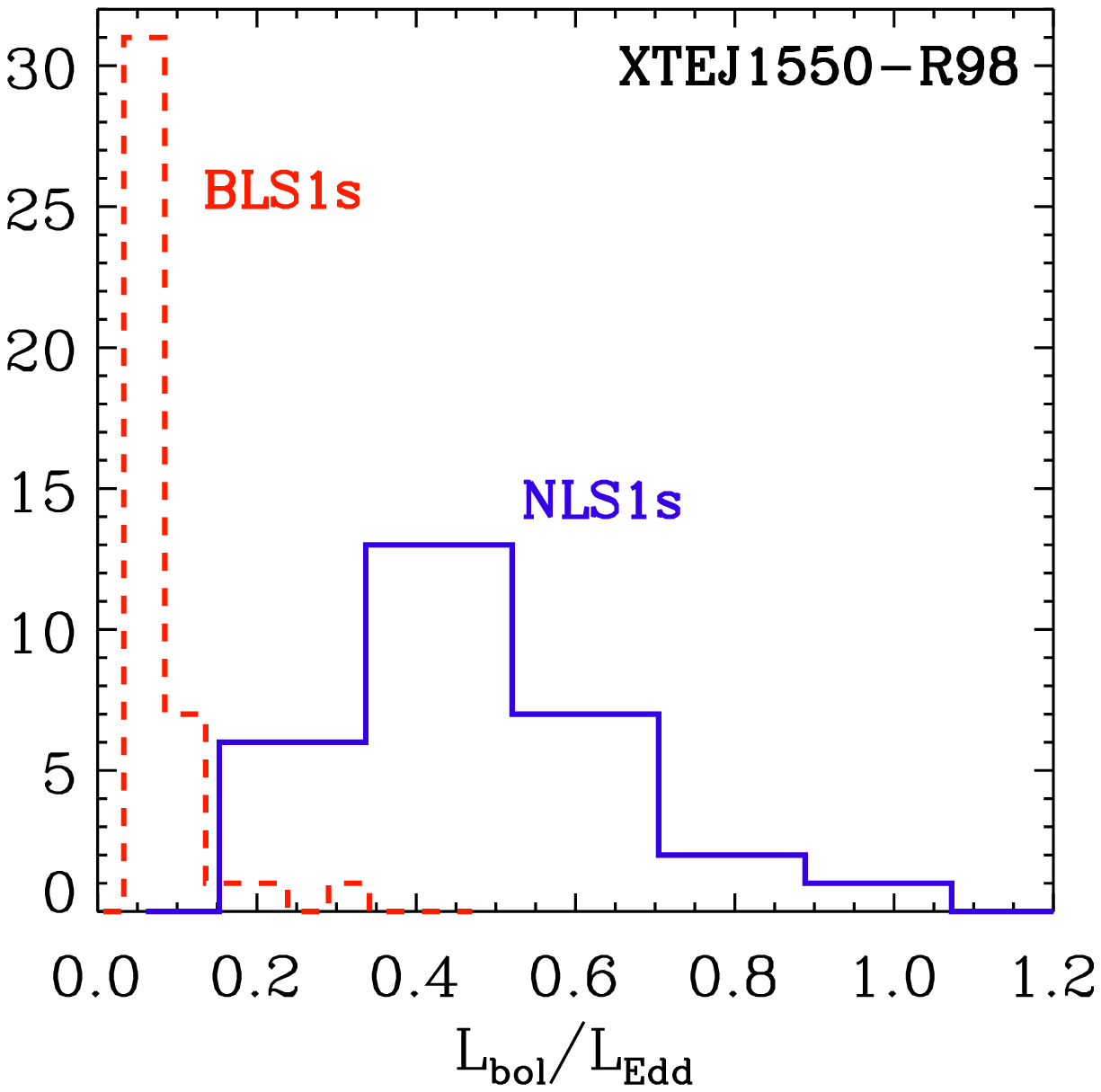}
	\hfill
	\includegraphics[width=0.69\columnwidth]{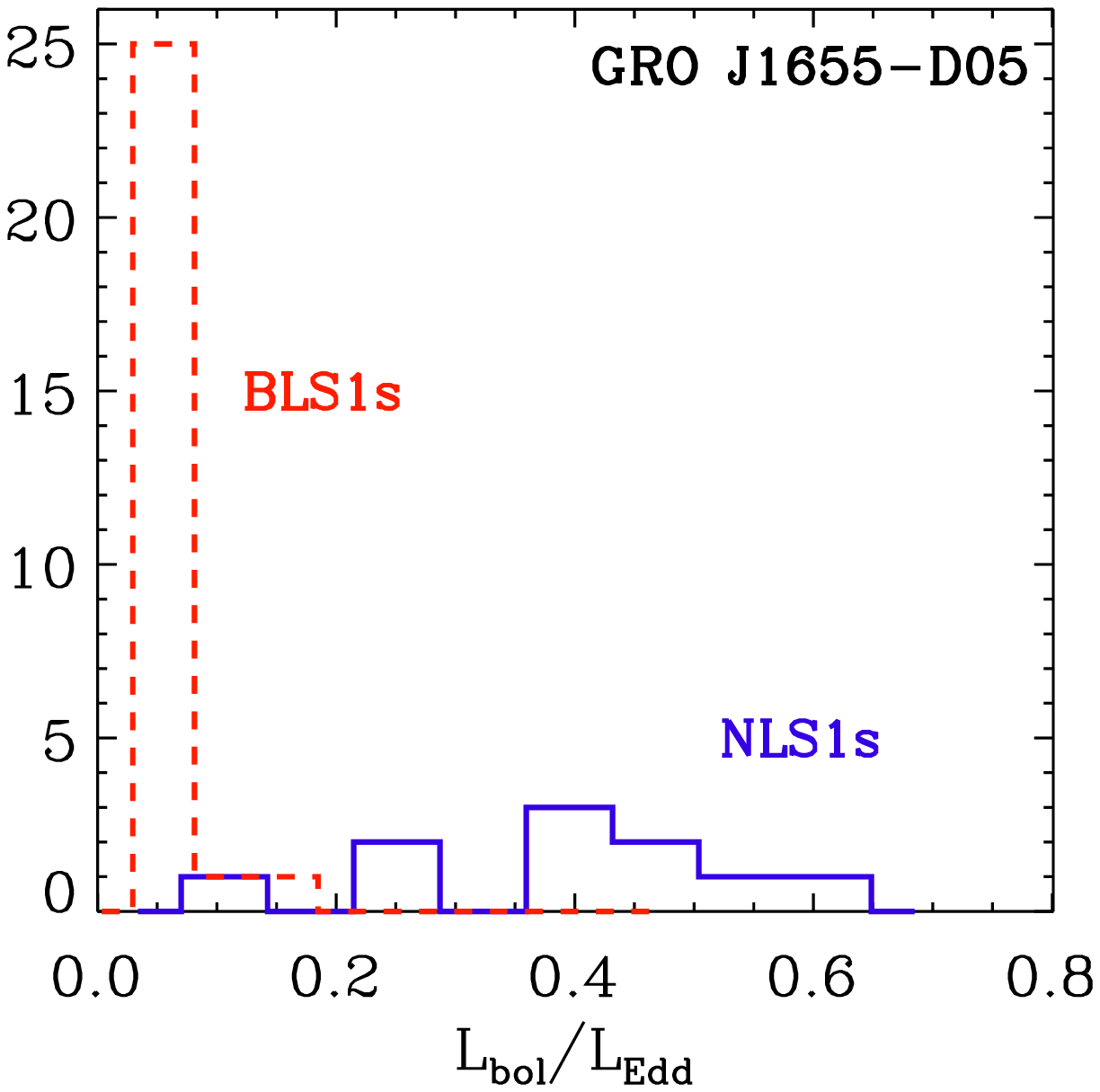}
	\hfill
	\includegraphics[width=0.69\columnwidth]{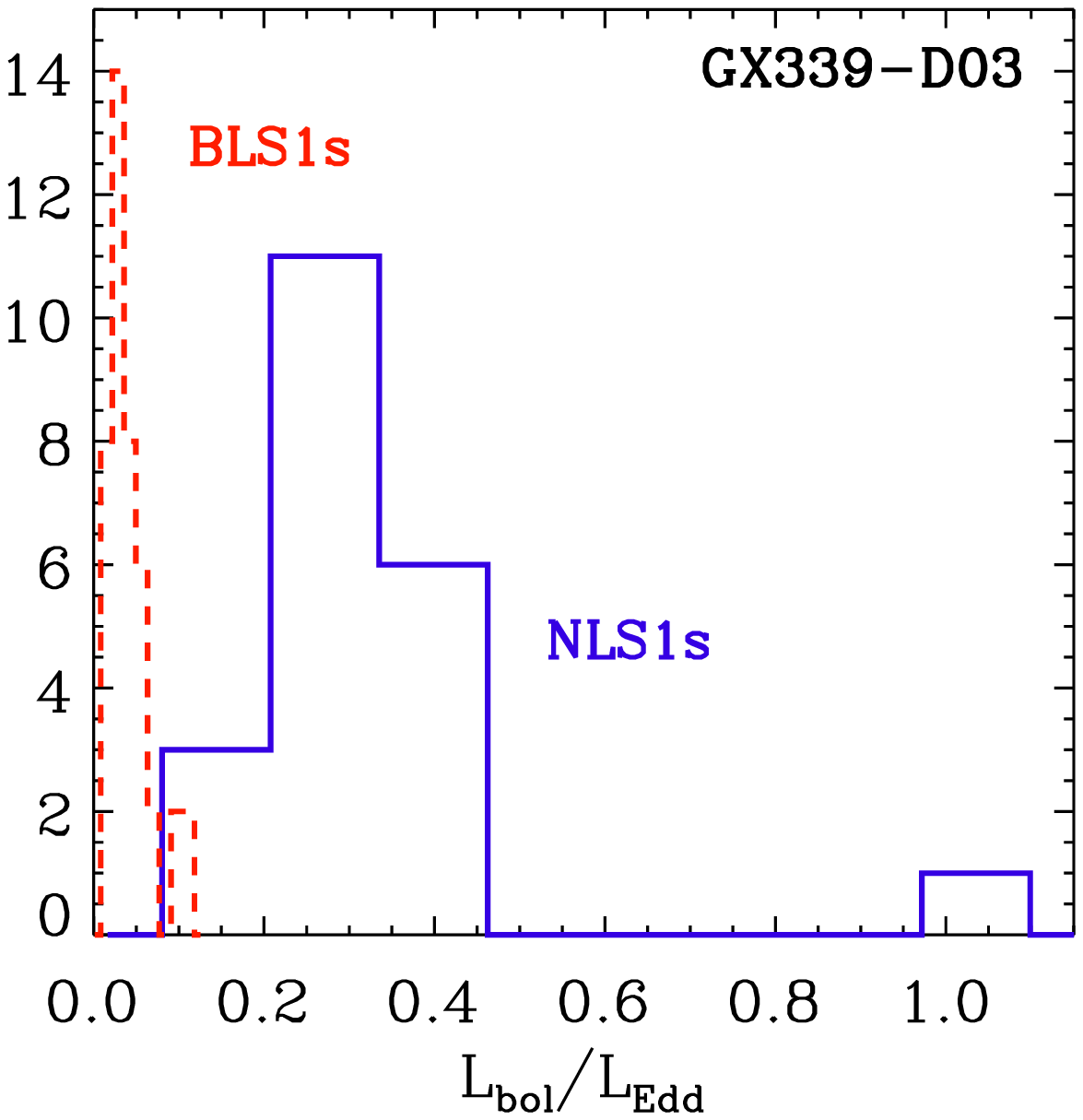}
	\caption{Number of NLS1s and BLS1s by \lbol/\ledd\ using reference source \xte\ (\textit{left}), \gro\ (\textit{center}), and \gx\ (\textit{right}).}
	\label{fig:histlbollEdd}
\end{figure*}

Our statistical analysis, based on a K--S test, reveals that the BH mass distributions of NLS1s and BLS1s, despite a substantial overlap (see Figure~\ref{fig:histmbh} and Table~\ref{tab:stats}), are inconsistent with the hypothesis that the two classes are drawn from the same population at the 99\% significance level. Similar results (that is, a clear distinction between NLS1s and BLS1s) with an even higher statistical significance are obtained when the accretion rate indicators (parameterized by \lx/\ledd\ and by \lbol/\ledd) of the NLS1 and BLS1 populations are compared (see Figures~\ref{fig:histlxlEdd} and \ref{fig:histlbollEdd}, and Table~\ref{tab:stats}).

One of the motivations of our work was to constrain \mbh\ of NLS1s and BLS1s with a method that, unlike the optically based techniques, does not make any assumptions about the BLR geometry and dynamics. It is therefore interesting to compare our results with the \mbh\ values constrained with the reverberation mapping method.  To this end, we used Georgia State University's AGN Black Hole Mass Database \citep{bent15}, which contains all \mbh\ values reported in peer-reviewed articles and has 24 objects in common with our samples.

\begin{figure*}
	\includegraphics[width=0.69\columnwidth]{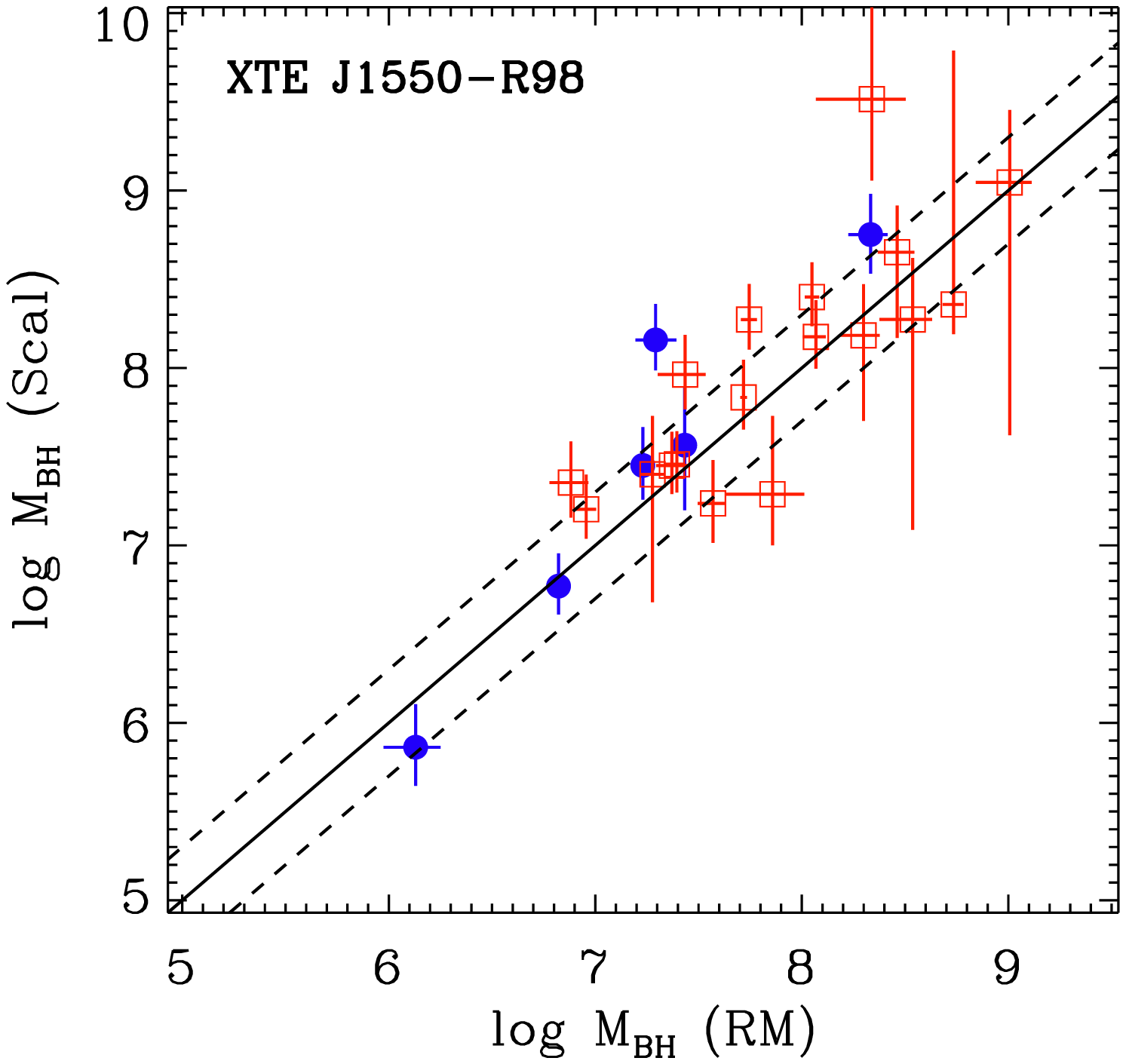}
	\hfill
	\includegraphics[width=0.69\columnwidth]{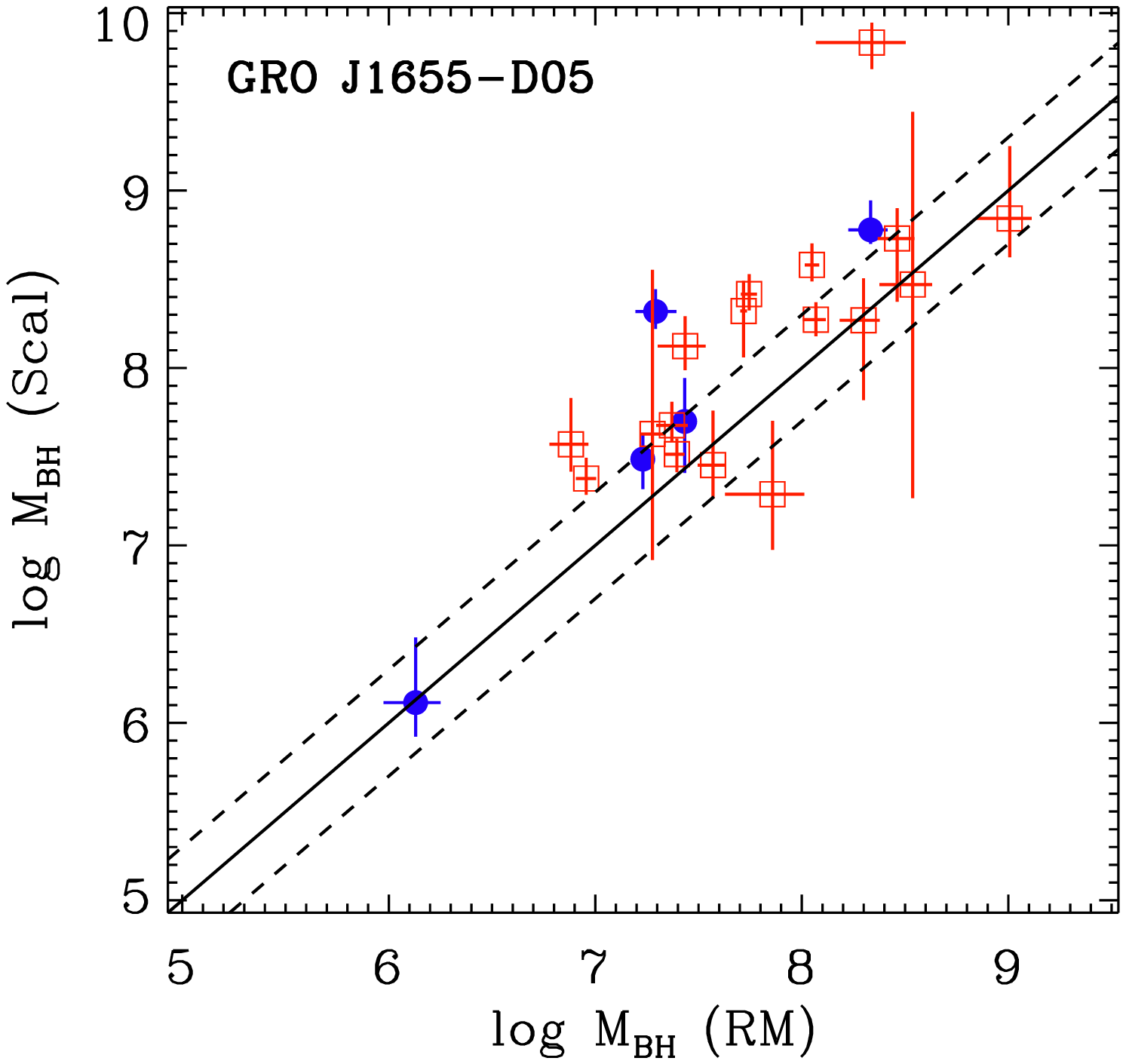}
	\hfill
	\includegraphics[width=0.69\columnwidth]{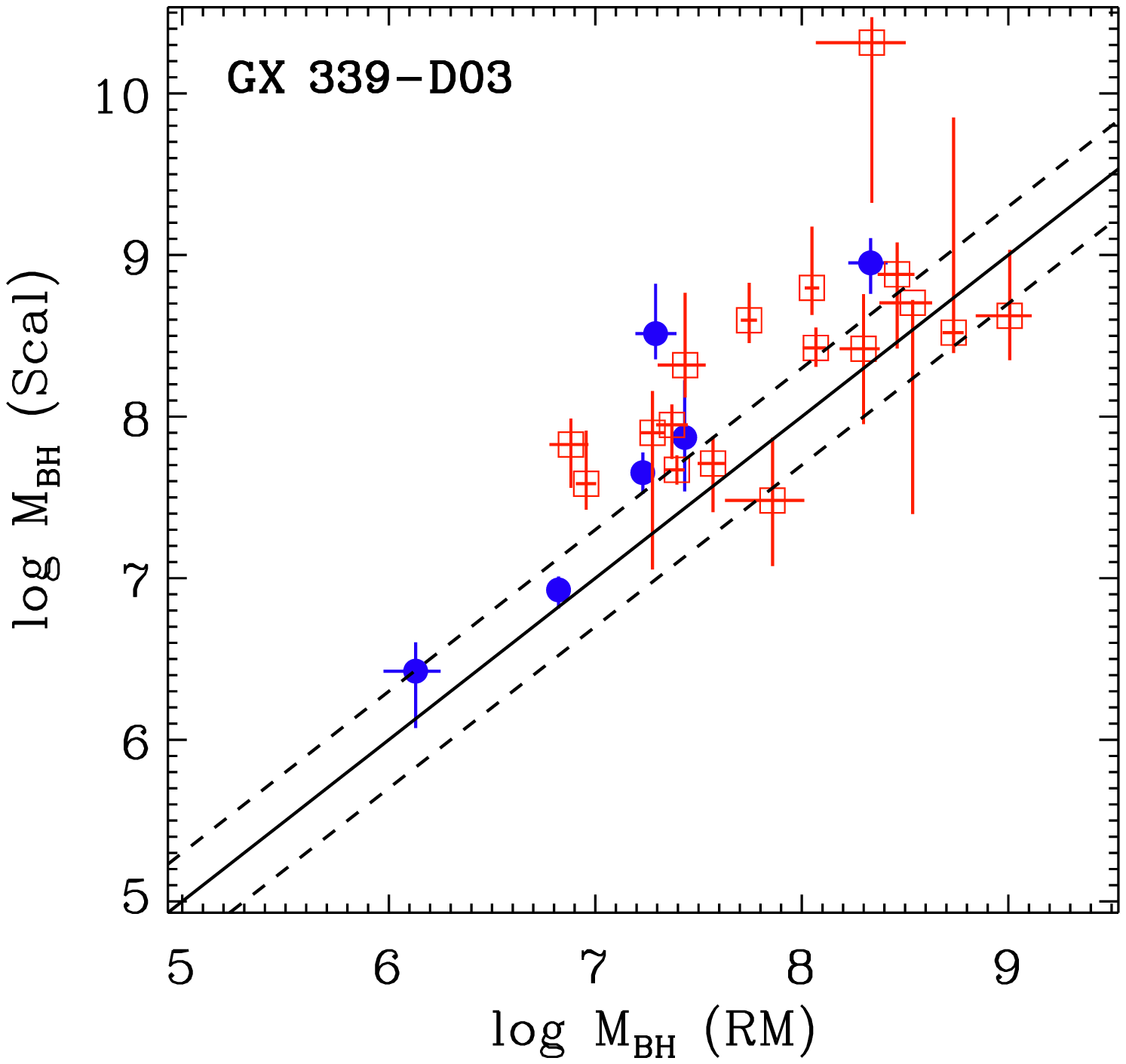}
	\caption{Comparison of reverberation mapping with the X-ray scaling method using reference source \xte\ (\textit{left}), \gro\ (\textit{center}), and \gx\ (\textit{right}). NLS1s are shown with filled-in circles (blue in the online version), and BLS1s are shown with open squares (red online). The y-axis shows log \mbh\ derived from the X-ray scaling method. The x-axis shows log \mbh\ derived from reverberation mapping. The dashed lines represent the 0.3 dex levels, commonly assumed as uncertainty on the reverberation mapping estimates.}
	\label{fig:correlation}
\end{figure*}

This comparison is illustrated in Figure~\ref{fig:correlation}, where the log of the \mbh\ values derived using the X-ray scaling method are plotted versus the corresponding values determined from reverberation mapping. The filled-in circles (blue in the online version) represent NLS1s, whereas the open squares (red online) represent BLS1s. A visual inspection of these plots suggests a broad agreement between the \mbh\ values obtained with the two methods for both NLS1s and BLS1s. From a closer look at these plots, it appears that NLS1 \mbh\ values obtained with the X-ray scaling method tend to lie above the one-to-one correlation line, whereas the BLS1 values seem to be more uniformly distributed around the one-to-one correlation. To compare the \mbh\ values obtained with these two methods in a more quantitative way, we computed the average ratios between the values determined with the scaling method and the reverberation method for NLS1s and BLS1s separately. Using the three different reference sources we obtained: $\langle M_{\rm BH,scal}/M_{\rm BH,RM}\rangle=1.03\pm0.02$ and $1.02\pm0.01$ for NLS1s and BLS1s, respectively, using \xte\ as reference, $1.05\pm0.02$ and $1.04\pm0.01$ using \gro, and $1.07\pm0.02$ and $1.06\pm0.02$ using \gx. The ratios of NLS1 values are statistically indistinguishable from the BLS1 values. This result, beside confirming the agreement between the X-ray scaling and reverberation mapping (RM), may suggest that the broad-line region dynamics is not strongly affected by radiation pressure and RM can be safely applied also to NLS1s in general. However, we must note that none of the NLS1s that are common to our sample and the RM one are accreting at or above the Eddington limit. We can therefore only conclude that the effect of the radiation pressure on the BLR dynamics is moderate and that RM can be safely used for this limited sample of moderately accreting NLS1s.

We also investigated whether our NLS1 sample can be considered homogeneous. To this end, we separated the NLS1s that have detected radio structures from those that have no radio detection (note that none of the objects with extended radio structure is classified as very radio-loud or has gamma-ray detections). We then compared these two subsamples of NLS1s: although the NLS1s with radio detections have slightly larger \mbh\ values, their average values are consistent within 1 sigma and the two distributions are indistinguishable based on a K--S test (with a probability of > 90\%).

Based on these results we can conclude that indeed NLS1s are characterized by different distributions of \mbh\ and accretion rates than `normal' BLS1s. This is in agreement with the hypothesis that NLS1s represent a younger phase of AGN activity, when relatively small supermassive black holes grow very rapidly.

However, before drawing any general conclusion one must consider that the sample used in this work is not complete by any means and may not be representative of the entire population of NLS1s. Indeed, the relatively high X-ray luminosity may indicate that the sample is skewed toward bright objects with large \mbh. Future studies, based on larger and volume-limited samples, will provide more conclusive results on the nature of NLS1s.

\begin{table*}
	\caption{NLS1 black hole masses}
	\begin{center}
		\begin{tabular}{lcccc} 
			\toprule
			\toprule       
			& \multicolumn{1}{c}{\xte} & \multicolumn{1}{c}{\gro} & \multicolumn{1}{c}{\gx} & \multicolumn{1}{c}{RM} \\
			\cmidrule(lr){2-2}\cmidrule(lr){3-3}\cmidrule(lr){4-4}\cmidrule(lr){5-5}
			Name & log \mbh & log \mbh & log \mbh & log \mbh \\
			(1) & (2) & (3) & (4) & (5) \\
			\midrule
			Mrk 335 & $7.45$ & $7.49$ & $7.65$ & $7.230$ (1,2,3) \\
			I Zw 1 & $7.91$ & \dots & \dots & \dots \\
			Ton S180 & $7.62$ & \dots & $7.75$ & \dots \\
			Mrk 359 & $6.93$ & $7.10$ & $7.30$ & \dots \\
			Mrk 1014 & $7.86$ & \dots & $8.01$ & \dots \\
			Mrk 586 & $7.94$ & \dots & \dots & \dots \\
			Mrk 1044 & $6.92$ & $6.92$ & $7.12$ & \dots \\
			RBS 416 & $7.25$ & $7.20$ & $7.43$ & \dots \\
			HE 0450-2958 & $8.59$ & \dots & $8.75$ & \dots \\
			PKS 0558-504 & $8.99$ & \dots & $9.17$ & \dots \\
			Mrk 110 & $8.16$ & $8.32$ & $8.51$ & $7.292$ (1,4) \\
			PG 0953+414 & $8.75$ & $8.78$ & $8.95$ & $8.333$ (1,5) \\
			RE J1034+396 & $7.00$ & \dots & $7.17$ & \dots \\
			PG 1115+407 & $8.00$ & \dots & $8.18$ & \dots \\
			PG 1116+215 & $8.50$ & $8.58$ & $8.73$ & \dots \\
			NGC 4051 & $5.86$ & $6.11$ & $6.42$ & $6.130$ (6) \\
			PG 1211+143 & $7.67$ & \dots & $7.83$ & \dots \\
			Mrk 766 & $6.77$ & \dots & $6.93$ & $6.822$ (7) \\
			Was 61 & $7.64$ & \dots & \dots & \dots \\
			PG 1244+026 & $7.38$ & \dots & \dots & \dots \\
			PG 1322+659 & $8.28$ & \dots & $8.31$ & \dots \\
			MCG-6-30-15 & $7.16$ & $7.21$ & $7.37$ & \dots \\
			IRAS 13349+2438 & $7.79$ & \dots & \dots & \dots \\
			PG 1402+261 & $8.30$ & $8.40$ & $8.56$ & \dots \\
			NGC 5506 & $7.02$ & $7.12$ & $7.28$ & \dots \\
			PG 1440+356 & $7.57$ & \dots & $7.75$ & \dots \\
			PG 1448+273 & $7.77$ & \dots & \dots & \dots \\
			Mrk 493 & $6.64$ & $6.59$ & $6.83$ & \dots \\
			IRAS 17020+4544 & $7.79$ & \dots & \dots & \dots \\
			PDS 456 & $8.40$ & \dots & $8.53$ & \dots \\
			Mrk 896 & $6.73$ & \dots & $6.90$ & \dots \\
			Mrk 1513 & $7.56$ & $7.70$ & $7.87$ & $7.433$ (1,2) \\
			II Zw 177 & $7.06$ & \dots & \dots & \dots \\
			Ark 564 & $7.37$ & \dots & \dots & \dots \\
			AM 2354-304 & $6.79$ & \dots & \dots & \dots \\
			\bottomrule
		\end{tabular}
	\end{center}
	\begin{flushleft}
		Columns: 1 = AGN name. 2 = log of BH mass scaled with reference source \xte. 3 = same for reference source \gro. 4 = same for reference source \gx. 5 = log of BH mass obtained with reverberation mapping (source: Georgia State University's AGN Black Hole Mass Database \citep{bent15}). References for RM values: (1) \citep[]{pete04}, (2) \citep[]{grie12}, (3) \citep[]{du14}, (4) \citep[]{koll01}, (5) \citep[]{kasp00}, (6) \citep[]{denn09}, (7) \citep[]{bent10}.
	\end{flushleft}
	\label{tab:nls1s}
\end{table*}  

\begin{table*}
	\caption{BLS1 black hole masses}
	\begin{center}
		\begin{tabular}{lcccc} 
			\toprule
			\toprule       
			& \multicolumn{1}{c}{\xte} & \multicolumn{1}{c}{\gro} & \multicolumn{1}{c}{\gx} & \multicolumn{1}{c}{RM} \\
			\cmidrule(lr){2-2}\cmidrule(lr){3-3}\cmidrule(lr){4-4}\cmidrule(lr){5-5}
			Name & log \mbh & log \mbh & log \mbh & log \mbh \\
			(1) & (2) & (3) & (4) & (5) \\
			\midrule
			PG 0052+251 & $8.65$ & $8.73$ & $8.88$ & $8.462$ (1,2) \\
			Q 0056-363 & $9.48$ & $9.58$ & $9.73$ & \dots \\
			Mrk 1152 & $7.76$ & $8.10$ & $8.68$ & \dots \\
			ESO 244-G17 & $6.78$ & $6.92$ & $7.10$ & \dots \\
			Fairall 9 & $8.18$ & $8.27$ & $8.42$ & $8.299$ (1,3,4) \\
			Mrk 590 & $7.24$ & $7.45$ & $7.71$ & $7.570$ (1) \\
			ESO 198-G24 & $7.93$ & $8.12$ & $8.36$ & \dots \\
			Fairall 1116 & $7.74$ & $7.82$ & $7.97$ & \dots \\
			1H 0419-577 & $8.82$ & $9.01$ & $9.25$ & \dots \\
			3C 120 & $8.27$ & $8.42$ & $8.60$ & $7.745$ (1,5,6) \\
			H 0439-272 & $8.20$ & $8.33$ & $8.50$ & \dots \\
			MCG-01-13-25 & $7.13$ & $7.37$ & $7.66$ & \dots \\
			Ark 120 & $8.18$ & $8.27$ & $8.43$ & $8.068$ (1,7) \\
			MCG-02-14-09 & $7.07$ & $7.18$ & $7.34$ & \dots \\
			MCG+8-11-11 & $7.90$ & $8.15$ & $8.46$ & \dots \\
			H 0557-385 & $8.13$ & $8.25$ & $8.42$ & \dots \\
			PMN J0623-6436 & $8.46$ & $8.63$ & $8.84$ & \dots \\
			ESO 209-G12 & $7.65$ & $7.77$ & $7.93$ & \dots \\
			PG 0804+761 & $8.36$ & \dots & $8.52$ & $8.735$ (1,2) \\
			Fairall 1146 & $7.61$ & $7.70$ & $7.85$ & \dots \\
			PG 0844+349 & $7.29$ & $7.29$ & $7.48$ & $7.858$ (1,2) \\
			MCG+04-22-42 & $7.83$ & $7.97$ & $8.15$ & \dots \\
			PG 0947+396 & $8.56$ & $8.72$ & $8.92$ & \dots \\
			HE 1029-1401 & $8.59$ & $8.66$ & $8.82$ & \dots \\
			PG 1048+342 & $8.17$ & $8.30$ & $8.47$ & \dots \\
			NGC 3516 & $7.46$ & $7.51$ & $7.67$ & $7.395$ (8) \\
			PG 1114+445 & $8.49$ & $8.63$ & $8.79$ & \dots \\
			NGC 3783 & $7.45$ & $7.68$ & $7.95$ & $7.371$ (9,10,11)\\
			HE 1143-1810 & $8.14$ & $8.29$ & $8.48$ & \dots \\
			PG 1202+281 & $8.72$ & $8.92$ & $9.15$ & \dots \\
			Mrk 205 & $8.10$ & \dots & $8.27$ & \dots \\
			Ark 374 & $7.80$ & $7.90$ & $8.05$ & \dots \\
			NGC 4593 & $7.35$ & $7.57$ & $7.83$ & $6.882$ (12,13) \\
			PG 1307+085 & $8.27$ & $8.47$ & $8.70$ & $8.537$ (1,2) \\
			4U 1344-60 & $6.64$ & $6.94$ & $7.35$ & \dots \\
			IC 4329A & $8.03$ & $8.19$ & $8.39$ & \dots \\
			Mrk 279 & $7.96$ & $8.12$ & $8.30$ & $7.435$ (1,14,15) \\
			PG 1352+183 & $9.25$ & \dots & \dots & \dots \\
			PG 1415+451 & $7.74$ & $7.79$ & $7.95$ & \dots \\
			NGC 5548 & $7.83$ & $8.32$ & \dots & $7.718$ (1,8,16,17,18,19,20,21) \\
			PG 1416-129 & $8.68$ & $8.98$ & $9.39$ & \dots \\
			PG 1425+267 & $9.10$ & $9.31$ & $9.57$ & \dots \\
			Mrk 1383 & $9.05$ & $8.84$ & $8.62$ & $9.007$ (1,2) \\
			PG 1427+480 & $8.63$ & $8.63$ & $8.82$ & \dots \\
			Mrk 841 & $8.58$ & $8.37$ & $8.16$ & \dots \\
			Mrk 290 & $7.40$ & $7.63$ & $7.90$ & $7.277$ (8) \\
			Mrk 876 & $9.51$ & $9.83$ & $10.31$ & $8.339$ (1,2) \\
			PG 1626+554 & $8.57$ & $8.89$ & $9.37$ & \dots \\
			IGR J17418-1212 & $7.80$ & $7.90$ & $8.06$ & \dots \\
			Mrk 509 & $8.40$ & $8.58$ & $8.80$ & $8.049$ (1) \\
			Mrk 304 & $7.50$ & $7.59$ & $7.74$ & \dots \\
			MR 2251-178 & $8.88$ & $9.15$ & $9.49$ & \dots \\
			NGC 7469 & $7.20$ & $7.38$ & $7.59$ & $6.956$ (22,23,24) \\
			Mrk 926 & $8.41$ & $8.60$ & $8.82$ & \dots \\
			\bottomrule
		\end{tabular}
	\end{center}
	\begin{flushleft}
		Columns: 1 = AGN name. 2 = log of BH mass scaled with reference source \xte. 3 = same for reference source \gro. 4 = same for reference source \gx. 5 = log of BH mass obtained with reverberation mapping (source: Georgia State University's AGN Black Hole Mass Database \citep{bent15}). References for RM values: (1) \citep[]{pete04}, (2) \citep[]{kasp00}, (3) \citep[]{sant97}, (4) \citep[]{rodr97}, (5) \citep[]{koll14}, (6) \citep[]{grie12}, (7) \citep[]{doro08}, (8) \citep[]{denn10}, (9) \citep[]{stir94}, (10) \citep[]{onke02}, (11) \citep[]{reic94}, (12) \citep[]{denn06}, (13) \citep[]{bart13}, (14) \citep[]{maoz90}, (15) \citep[]{sant01}, (16) \citep[]{netz90}, (17) \citep[]{kova14}, (18) \citep[]{bent10}, (19) \citep[]{diet93}, (20) \citep[]{clav91}, (21) \citep[]{kori95}, (22) \citep[]{coll98}, (23) \citep[]{pete14}, (24) \citep[]{wand97}.
	\end{flushleft}
	\label{tab:bls1s}
\end{table*}  

\begin{table*}
	\caption{NLS1 black hole accretion rates}
	\begin{center}
		\begin{tabular}{lcccccc} 
			\toprule
			\toprule       
			& \multicolumn{2}{c}{\xte} & \multicolumn{2}{c}{\gro} & \multicolumn{2}{c}{\gx} \\
			\cmidrule(lr){2-3}\cmidrule(lr){4-5}\cmidrule(lr){6-7}
			Name & \lx/\ledd & \lbol/\ledd & \lx/\ledd & \lbol/\ledd & \lx/\ledd & \lbol/\ledd \\
			(1) & (2) & (3) & (4) & (5) & (6) & (7) \\
			\midrule
			Mrk 335 & $2.59 \times 10^{-3}$ & $5.10 \times 10^{-1}$ & $2.39 \times 10^{-3}$ & $4.70 \times 10^{-1}$ & $1.62 \times 10^{-3}$ & $3.20 \times 10^{-1}$ \\
			I Zw 1 & $6.45 \times 10^{-3}$ & $6.21 \times 10^{-1}$ & \dots & \dots & \dots & \dots \\
			Ton S180 & $6.51 \times 10^{-3}$ & $6.28 \times 10^{-1}$ & \dots & \dots & $4.73 \times 10^{-3}$ & $4.56 \times 10^{-1}$ \\
			Mrk 359 & $4.22 \times 10^{-3}$ & $4.07 \times 10^{-1}$ & $2.87 \times 10^{-3}$ & $2.77 \times 10^{-1}$ & $1.81 \times 10^{-3}$ & $1.74 \times 10^{-1}$ \\
			Mrk 1014 & $6.82 \times 10^{-3}$ & $6.58 \times 10^{-1}$ & \dots & \dots & $4.76 \times 10^{-3}$ & $4.59 \times 10^{-1}$ \\
			Mrk 586 & $8.88 \times 10^{-3}$ & $8.56 \times 10^{-1}$ & \dots & \dots & \dots & \dots \\
			Mrk 1044 & $5.86 \times 10^{-3}$ & $5.65 \times 10^{-1}$ & $5.86 \times 10^{-3}$ & $5.65 \times 10^{-1}$ & $3.76 \times 10^{-3}$ & $3.62 \times 10^{-1}$ \\
			RBS 416 & $6.95 \times 10^{-3}$ & $6.70 \times 10^{-1}$ & $7.85 \times 10^{-3}$ & $7.57 \times 10^{-1}$ & $4.53 \times 10^{-3}$ & $4.37 \times 10^{-1}$ \\
			HE 0450-2958 & $7.35 \times 10^{-3}$ & $7.09 \times 10^{-1}$ & \dots & \dots & $5.07 \times 10^{-3}$ & $4.89 \times 10^{-1}$ \\
			PKS 0558-504 & $5.95 \times 10^{-3}$ & $5.74 \times 10^{-1}$ & \dots & \dots & $3.91 \times 10^{-3}$ & $3.77 \times 10^{-1}$ \\
			Mrk 110 & $4.38 \times 10^{-3}$ & $8.07 \times 10^{-2}$ & $3.03 \times 10^{-3}$ & $5.57 \times 10^{-2}$ & $1.93 \times 10^{-3}$ & $3.56 \times 10^{-2}$ \\
			PG 0953+414 & $6.40 \times 10^{-3}$ & $6.17 \times 10^{-1}$ & $6.03 \times 10^{-3}$ & $5.81 \times 10^{-1}$ & $4.04 \times 10^{-3}$ & $3.90 \times 10^{-1}$ \\
			RE J1034+396 & $2.73 \times 10^{-3}$ & $2.64 \times 10^{-1}$ & \dots & \dots & $1.85 \times 10^{-3}$ & $1.78 \times 10^{-1}$ \\
			PG 1115+407 & $1.97 \times 10^{-2}$ & $1.90$ & \dots & \dots & $1.30 \times 10^{-2}$ & $1.25$ \\
			PG 1116+215 & $6.30 \times 10^{-3}$ & $6.07 \times 10^{-1}$ & $5.22 \times 10^{-3}$ & $5.04 \times 10^{-1}$ & $3.69 \times 10^{-3}$ & $3.56 \times 10^{-1}$ \\
			NGC 4051 & $1.48 \times 10^{-3}$ & $6.08 \times 10^{-2}$ & $8.28 \times 10^{-4}$ & $3.40 \times 10^{-2}$ & $4.05 \times 10^{-4}$ & $1.67 \times 10^{-2}$ \\
			PG 1211+143 & $8.55 \times 10^{-3}$ & $1.85$ & \dots & \dots & $5.96 \times 10^{-3}$ & $1.29$ \\
			Mrk 766 & $7.43 \times 10^{-3}$ & $7.17 \times 10^{-1}$ & \dots & \dots & $5.18 \times 10^{-3}$ & $5.00 \times 10^{-1}$ \\
			Was 61 & $4.62 \times 10^{-3}$ & $4.46 \times 10^{-1}$ & \dots & \dots & \dots & \dots \\
			PG 1244+026 & $4.49 \times 10^{-3}$ & $4.33 \times 10^{-1}$ & \dots & \dots & \dots & \dots \\
			PG 1322+659 & $4.79 \times 10^{-3}$ & $4.62 \times 10^{-1}$ & \dots & \dots & $4.54 \times 10^{-3}$ & $4.37 \times 10^{-1}$ \\
			MCG-6-30-15 & $2.19 \times 10^{-3}$ & $2.11 \times 10^{-1}$ & $1.95 \times 10^{-3}$ & $1.88 \times 10^{-1}$ & $1.35 \times 10^{-3}$ & $1.30 \times 10^{-1}$ \\
			IRAS 13349+2438 & $1.20 \times 10^{-2}$ & $1.16$ & \dots & \dots & \dots & \dots \\
			PG 1402+261 & $4.28 \times 10^{-3}$ & $4.13 \times 10^{-1}$ & $3.34 \times 10^{-3}$ & $3.22 \times 10^{-1}$ & $2.33 \times 10^{-3}$ & $2.24 \times 10^{-1}$ \\
			NGC 5506 & $8.10 \times 10^{-3}$ & $7.81 \times 10^{-1}$ & $6.37 \times 10^{-3}$ & $6.14 \times 10^{-1}$ & $4.46 \times 10^{-3}$ & $4.30 \times 10^{-1}$ \\
			PG 1440+356 & $5.57 \times 10^{-3}$ & $5.37 \times 10^{-1}$ & \dots & \dots & $3.67 \times 10^{-3}$ & $3.53 \times 10^{-1}$ \\
			PG 1448+273 & $3.62 \times 10^{-3}$ & $3.49 \times 10^{-1}$ & \dots & \dots & \dots & \dots \\
			Mrk 493 & $6.87 \times 10^{-3}$ & $6.62 \times 10^{-1}$ & $7.77 \times 10^{-3}$ & $7.49 \times 10^{-1}$ & $4.48 \times 10^{-3}$ & $4.32 \times 10^{-1}$ \\
			IRAS 17020+4544 & $6.71 \times 10^{-3}$ & $6.47 \times 10^{-1}$ & \dots & \dots & \dots & \dots \\
			PDS 456 & $7.91 \times 10^{-3}$ & $7.62 \times 10^{-1}$ & \dots & \dots & $5.85 \times 10^{-3}$ & $5.64 \times 10^{-1}$ \\
			Mrk 896 & $8.31 \times 10^{-3}$ & $8.01 \times 10^{-1}$ & \dots & \dots & $5.67 \times 10^{-3}$ & $5.46 \times 10^{-1}$ \\
			Mrk 1513 & $6.91 \times 10^{-3}$ & $6.67 \times 10^{-1}$ & $5.08 \times 10^{-3}$ & $4.90 \times 10^{-1}$ & $3.42 \times 10^{-3}$ & $3.29 \times 10^{-1}$ \\
			II Zw 177 & $1.07 \times 10^{-2}$ & $1.03$ & \dots & \dots & \dots & \dots \\
			Ark 564 & $9.90 \times 10^{-3}$ & $9.55 \times 10^{-1}$ & \dots & \dots & \dots & \dots \\
			AM 2354-304 & $8.52 \times 10^{-3}$ & $8.21 \times 10^{-1}$ & \dots & \dots & \dots & \dots \\
			\bottomrule
		\end{tabular}
	\end{center}
	\begin{flushleft}
		Columns: 1 = AGN name. 2 = ratio of X-ray luminosity (2--10 keV) to Eddington luminosity scaled with \xte. 3 = ratio of bolometric luminosity to Eddington luminosity scaled with \xte. 4--5 = same for reference source \gro. 6--7 = same for reference source \gx.
	\end{flushleft}
	\label{tab:nls1s_ar}
\end{table*}  

\begin{table*}
	\caption{BLS1 black hole accretion rates}
	\footnotesize
	\begin{center}
		\begin{tabular}{lcccccc} 
	\toprule
	\toprule       
	& \multicolumn{2}{c}{\xte} & \multicolumn{2}{c}{\gro} & \multicolumn{2}{c}{\gx} \\
	\cmidrule(lr){2-3}\cmidrule(lr){4-5}\cmidrule(lr){6-7}
	Name & \lx/\ledd & \lbol/\ledd & \lx/\ledd & \lbol/\ledd & \lx/\ledd & \lbol/\ledd \\
	(1) & (2) & (3) & (4) & (5) & (6) & (7) \\
	\midrule
			PG 0052+251 & $5.82 \times 10^{-3}$ & $1.14 \times 10^{-1}$ & $4.87 \times 10^{-3}$ & $9.50 \times 10^{-2}$ & $3.44 \times 10^{-3}$ & $6.72 \times 10^{-2}$ \\
			Q 0056-363 & $4.80 \times 10^{-4}$ & $1.14 \times 10^{-2}$ & $3.84 \times 10^{-4}$ & $9.09 \times 10^{-3}$ & $2.70 \times 10^{-4}$ & $6.40 \times 10^{-3}$ \\
			Mrk 1152 & $4.04 \times 10^{-3}$ & $9.60 \times 10^{-2}$ & $1.84 \times 10^{-3}$ & $4.36 \times 10^{-2}$ & $4.79 \times 10^{-4}$ & $1.14 \times 10^{-2}$ \\
			ESO 244-G17 & $4.59 \times 10^{-3}$ & $1.09 \times 10^{-1}$ & $3.33 \times 10^{-3}$ & $7.92 \times 10^{-2}$ & $2.22 \times 10^{-3}$ & $5.28 \times 10^{-2}$ \\
			Fairall 9 & $5.54 \times 10^{-3}$ & $5.82 \times 10^{-2}$ & $4.55 \times 10^{-3}$ & $4.78 \times 10^{-2}$ & $3.22 \times 10^{-3}$ & $3.38 \times 10^{-2}$ \\
			Mrk 590 & $4.46 \times 10^{-3}$ & $3.12 \times 10^{-2}$ & $2.71 \times 10^{-3}$ & $1.90 \times 10^{-2}$ & $1.50 \times 10^{-3}$ & $1.05 \times 10^{-2}$ \\
			ESO 198-G24 & $4.45 \times 10^{-3}$ & $1.06 \times 10^{-1}$ & $2.83 \times 10^{-3}$ & $6.71 \times 10^{-2}$ & $1.65 \times 10^{-3}$ & $3.92 \times 10^{-2}$ \\
			Fairall 1116 & $5.93 \times 10^{-3}$ & $1.41 \times 10^{-1}$ & $4.87 \times 10^{-3}$ & $1.16 \times 10^{-2}$ & $3.44 \times 10^{-3}$ & $8.17 \times 10^{-2}$ \\
			1H 0419-577 & $3.99 \times 10^{-3}$ & $9.48 \times 10^{-2}$ & $2.54 \times 10^{-3}$ & $6.02 \times 10^{-2}$ & $1.48 \times 10^{-3}$ & $3.52 \times 10^{-2}$ \\
			3C 120 & $4.52 \times 10^{-3}$ & $9.30 \times 10^{-2}$ & $3.24 \times 10^{-3}$ & $6.68 \times 10^{-2}$ & $2.14 \times 10^{-3}$ & $4.41 \times 10^{-2}$ \\
			H 0439-272 & $4.67 \times 10^{-3}$ & $1.11 \times 10^{-1}$ & $3.47 \times 10^{-3}$ & $8.25 \times 10^{-2}$ & $2.35 \times 10^{-3}$ & $5.59 \times 10^{-2}$ \\
			MCG-01-13-25 & $3.96 \times 10^{-3}$ & $9.41 \times 10^{-2}$ & $2.29 \times 10^{-3}$ & $5.43 \times 10^{-2}$ & $1.17 \times 10^{-3}$ & $2.79 \times 10^{-2}$ \\
			Ark 120 & $4.52 \times 10^{-3}$ & $1.07 \times 10^{-1}$ & $3.61 \times 10^{-3}$ & $8.58 \times 10^{-2}$ & $2.54 \times 10^{-3}$ & $6.04 \times 10^{-2}$ \\
			MCG-02-14-09 & $5.07 \times 10^{-3}$ & $1.20 \times 10^{-1}$ & $3.88 \times 10^{-3}$ & $9.20 \times 10^{-2}$ & $2.68 \times 10^{-3}$ & $6.36 \times 10^{-2}$ \\
			MCG+8-11-11 & $4.09 \times 10^{-3}$ & $9.71 \times 10^{-2}$ & $2.29 \times 10^{-3}$ & $5.44 \times 10^{-2}$ & $1.12 \times 10^{-3}$ & $2.66 \times 10^{-2}$ \\
			H 0557-385 & $6.30 \times 10^{-3}$ & $1.50 \times 10^{-1}$ & $4.72 \times 10^{-3}$ & $1.12 \times 10^{-1}$ & $3.22 \times 10^{-3}$ & $7.65 \times 10^{-2}$ \\
			PMN J0623-6436 & $4.03 \times 10^{-3}$ & $9.58 \times 10^{-2}$ & $2.71 \times 10^{-3}$ & $6.41 \times 10^{-2}$ & $1.68 \times 10^{-3}$ & $3.97 \times 10^{-2}$ \\
			ESO 209-G12 & $5.36 \times 10^{-3}$ & $1.27 \times 10^{-1}$ & $4.10 \times 10^{-3}$ & $9.74 \times 10^{-2}$ & $2.83 \times 10^{-3}$ & $6.73 \times 10^{-2}$ \\
			PG 0804+761 & $6.17 \times 10^{-4}$ & $1.47 \times 10^{-2}$ & \dots & \dots & $4.26 \times 10^{-4}$ & $1.01 \times 10^{-2}$ \\
			Fairall 1146 & $5.62 \times 10^{-3}$ & $1.34 \times 10^{-1}$ & $4.58 \times 10^{-3}$ & $1.09 \times 10^{-1}$ & $3.23 \times 10^{-3}$ & $7.68 \times 10^{-2}$ \\
			PG 0844+349 & $7.24 \times 10^{-3}$ & $5.21 \times 10^{-1}$ & $7.24 \times 10^{-3}$ & $5.21 \times 10^{-1}$ & $4.65 \times 10^{-3}$ & $1.10 \times 10^{-1}$ \\
			MCG+04-22-42 & $4.35 \times 10^{-3}$ & $1.03 \times 10^{-1}$ & $3.13 \times 10^{-3}$ & $7.42 \times 10^{-2}$ & $2.06 \times 10^{-3}$ & $4.90 \times 10^{-2}$ \\
			PG 0947+396 & $4.35 \times 10^{-3}$ & $1.03 \times 10^{-1}$ & $3.01 \times 10^{-3}$ & $7.13 \times 10^{-2}$ & $1.92 \times 10^{-3}$ & $1.38 \times 10^{-1}$ \\
			HE 1029-1401 & $6.36 \times 10^{-3}$ & $1.51 \times 10^{-1}$ & $5.33 \times 10^{-3}$ & $1.26 \times 10^{-1}$ & $3.76 \times 10^{-3}$ & $8.94 \times 10^{-2}$ \\
			PG 1048+342 & $5.17 \times 10^{-3}$ & $1.23 \times 10^{-1}$ & $3.84 \times 10^{-3}$ & $9.10 \times 10^{-2}$ & $2.60 \times 10^{-3}$ & $6.16 \times 10^{-2}$ \\
			NGC 3516 & $2.07 \times 10^{-3}$ & $3.16 \times 10^{-2}$ & $1.81 \times 10^{-3}$ & $2.77 \times 10^{-2}$ & $1.27 \times 10^{-3}$ & $2.00 \times 10^{-2}$ \\
			PG 1114+445 & $2.96 \times 10^{-3}$ & $7.02 \times 10^{-2}$ & $2.17 \times 10^{-3}$ & $5.15 \times 10^{-2}$ & $1.50 \times 10^{-3}$ & $3.57 \times 10^{-2}$ \\
			NGC 3783 & $3.27 \times 10^{-3}$ & $5.17 \times 10^{-2}$ & $1.94 \times 10^{-3}$ & $3.07 \times 10^{-2}$ & $1.04 \times 10^{-3}$ & $8.00 \times 10^{-3}$ \\
			HE 1143-1810 & $3.94 \times 10^{-3}$ & $9.35 \times 10^{-2}$ & $2.76 \times 10^{-3}$ & $6.56 \times 10^{-2}$ & $1.78 \times 10^{-3}$ & $4.24 \times 10^{-2}$ \\
			PG 1202+281 & $3.78 \times 10^{-3}$ & $8.96 \times 10^{-2}$ & $2.40 \times 10^{-3}$ & $5.68 \times 10^{-2}$ & $1.40 \times 10^{-3}$ & $3.32 \times 10^{-2}$ \\
			Mrk 205 & $8.49 \times 10^{-3}$ & $2.02 \times 10^{-1}$ & \dots & \dots & $5.79 \times 10^{-3}$ & $1.38 \times 10^{-1}$ \\
			Ark 374 & $4.00 \times 10^{-3}$ & $9.50 \times 10^{-2}$ & $3.23 \times 10^{-3}$ & $7.67 \times 10^{-2}$ & $2.28 \times 10^{-3}$ & $5.41 \times 10^{-2}$ \\
			NGC 4593 & $1.50 \times 10^{-3}$ & $1.15 \times 10^{-2}$ & $9.11 \times 10^{-4}$ & $7.01 \times 10^{-3}$ & $5.04 \times 10^{-3}$ & $2.12 \times 10^{-2}$ \\
			PG 1307+085 & $4.55 \times 10^{-3}$ & $1.59 \times 10^{-1}$ & $2.89 \times 10^{-3}$ & $1.01 \times 10^{-1}$ & $1.69 \times 10^{-3}$ & $4.00 \times 10^{-2}$ \\
			4U 1344-60 & $3.76 \times 10^{-3}$ & $8.94 \times 10^{-2}$ & $1.89 \times 10^{-3}$ & $4.48 \times 10^{-2}$ & $7.26 \times 10^{-4}$ & $1.72 \times 10^{-2}$ \\
			IC 4329A & $4.13 \times 10^{-3}$ & $9.80 \times 10^{-2}$ & $2.81 \times 10^{-3}$ & $6.68 \times 10^{-2}$ & $1.77 \times 10^{-3}$ & $4.20 \times 10^{-2}$ \\
			Mrk 279 & $4.27 \times 10^{-3}$ & $8.67 \times 10^{-2}$ & $2.95 \times 10^{-3}$ & $6.00 \times 10^{-2}$ & $1.95 \times 10^{-3}$ & $3.97 \times 10^{-2}$ \\
			PG 1352+183 & $1.30 \times 10^{-3}$ & $3.08 \times 10^{-2}$ & \dots & \dots & \dots & \dots \\
			PG 1415+451 & $5.07 \times 10^{-3}$ & $1.21 \times 10^{-1}$ & $4.45 \times 10^{-3}$ & $1.05 \times 10^{-1}$ & $3.11 \times 10^{-3}$ & $1.09 \times 10^{-1}$ \\
			NGC 5548 & $2.14 \times 10^{-3}$ & $2.16 \times 10^{-2}$ & $6.98 \times 10^{-4}$ & $5.37 \times 10^{-3}$ & \dots & \dots \\
			PG 1416-129 & $3.06 \times 10^{-4}$ & $7.27 \times 10^{-3}$ & $1.53 \times 10^{-4}$ & $3.64 \times 10^{-3}$ & $5.90 \times 10^{-5}$ & $1.40 \times 10^{-3}$ \\
			PG 1425+267 & $3.69 \times 10^{-3}$ & $8.77 \times 10^{-2}$ & $2.25 \times 10^{-3}$ & $5.32 \times 10^{-2}$ & $1.24 \times 10^{-3}$ & $2.94 \times 10^{-2}$ \\
			Mrk 1383 & $1.06 \times 10^{-3}$ & $2.51 \times 10^{-2}$ & $1.69 \times 10^{-3}$ & $4.02 \times 10^{-2}$ & $2.80 \times 10^{-3}$ & $6.63 \times 10^{-2}$ \\
			PG 1427+480 & $3.80 \times 10^{-3}$ & $9.02 \times 10^{-2}$ & $3.80 \times 10^{-3}$ & $9.01 \times 10^{-2}$ & $2.44 \times 10^{-3}$ & $5.78 \times 10^{-2}$ \\
			Mrk 841 & $9.38 \times 10^{-4}$ & $2.23 \times 10^{-2}$ & $1.50 \times 10^{-3}$ & $3.55 \times 10^{-2}$ & $2.48 \times 10^{-3}$ & $5.87 \times 10^{-2}$ \\
			Mrk 290 & $4.57 \times 10^{-3}$ & $1.09 \times 10^{-1}$ & $2.71 \times 10^{-3}$ & $6.42 \times 10^{-2}$ & $1.45 \times 10^{-3}$ & $3.43 \times 10^{-2}$ \\
			Mrk 876 & $8.70 \times 10^{-4}$ & $2.07 \times 10^{-2}$ & $4.16 \times 10^{-4}$ & $9.87 \times 10^{-3}$ & $1.38 \times 10^{-4}$ & $3.27 \times 10^{-3}$ \\
			PG 1626+554 & $2.91 \times 10^{-3}$ & $6.92 \times 10^{-2}$ & $1.39 \times 10^{-3}$ & $3.31 \times 10^{-2}$ & $4.63 \times 10^{-4}$ & $1.10 \times 10^{-2}$ \\
			IGR J17418-1212 & $5.10 \times 10^{-3}$ & $1.21 \times 10^{-1}$ & $4.04 \times 10^{-3}$ & $9.60 \times 10^{-2}$ & $2.84 \times 10^{-3}$ & $6.74 \times 10^{-2}$ \\
			Mrk 509 & $4.29 \times 10^{-3}$ & $6.95 \times 10^{-2}$ & $2.83 \times 10^{-3}$ & $4.58 \times 10^{-2}$ & $1.72 \times 10^{-3}$ & $2.79 \times 10^{-2}$ \\
			Mrk 304 & $1.03 \times 10^{-2}$ & $2.44 \times 10^{-1}$ & $8.29 \times 10^{-3}$ & $1.96 \times 10^{-1}$ & $5.85 \times 10^{-3}$ & $1.39 \times 10^{-1}$ \\
			MR 2251-178 & $4.31 \times 10^{-3}$ & $1.02 \times 10^{-1}$ & $2.34 \times 10^{-3}$ & $5.55 \times 10^{-2}$ & $1.08 \times 10^{-3}$ & $2.55 \times 10^{-2}$ \\
			NGC 7469 & $8.66 \times 10^{-3}$ & $3.64 \times 10^{-1}$ & $5.81 \times 10^{-3}$ & $4.48 \times 10^{-2}$ & $3.60 \times 10^{-3}$ & $5.51 \times 10^{-2}$ \\
			Mrk 926 & $4.52 \times 10^{-3}$ & $1.07 \times 10^{-1}$ & $2.93 \times 10^{-3}$ & $6.94 \times 10^{-2}$ & $1.75 \times 10^{-3}$ & $4.14 \times 10^{-2}$ \\
			\bottomrule
		\end{tabular}
	\end{center}
	\begin{flushleft}
		Columns: 1 = AGN name. 2 = ratio of X-ray luminosity (2--10 keV) to Eddington luminosity scaled with \xte. 3 = ratio of bolometric luminosity to Eddington luminosity scaled with \xte. 4--5 = same for reference source \gro. 6--7 = same for reference source \gx.
	\end{flushleft}
	\label{tab:bls1s_ar}
\end{table*}

\begin{table*}
	\caption{Statistical comparison of NLS1s and BLS1s}	
	\begin{tabular}{lccccccccc} 
		\toprule
		\toprule       
		& \multicolumn{3}{c}{\xte} & \multicolumn{3}{c}{\gro} & \multicolumn{3}{c}{\gx} \\
		\cmidrule(lr){2-4}\cmidrule(lr){5-7}\cmidrule(lr){8-10}
		& \mbh & \lx/\ledd & \lbol/\ledd
		& \mbh & \lx/\ledd & \lbol/\ledd
		& \mbh & \lx/\ledd & \lbol/\ledd \\
		& ($\times 10^8$ \msun) & ($\times 10^{-3}$) & ($\times 10^{-1}$)
		& ($\times 10^8$ \msun) & ($\times 10^{-3}$) & ($\times 10^{-1}$)
		& ($\times 10^8$ \msun) & ($\times 10^{-3}$) & ($\times 10^{-1}$) \\
		\midrule
		NLS1s
		& $1.1$ & $6.7$ & $6.7$ 
		& $1.2$ & $4.5$ & $4.3$ 
		& $2.2$ & $4.1$ & $4.2$ \\
		BLS1s
		& $3.5$ & $4.2$ & $1.0$
		& $5.0$ & $3.1$ & $0.7$
		& $10$ & $2.1$ & $0.5$ \\
		\midrule
		Prob (K--S)
		& $1.0 \times 10^{-2}$ & $< 10^{-4}$ & $< 10^{-4}$
		& $1.1 \times 10^{-2}$ & $1.6 \times 10^{-2}$ & $< 10^{-4}$
		& $8.0 \times 10^{-3}$ & $< 10^{-4}$ & $< 10^{-4}$ \\
		Prob (t)
		& $3.6 \times 10^{-2}$ & $< 10^{-4}$ & $< 10^{-4}$
		& $2.2 \times 10^{-1}$ & $1.0 \times 10^{-2}$ & $1.0 \times 10^{-4}$
		& $1.8 \times 10^{-1}$ & $< 10^{-4}$ & $< 10^{-4}$ \\
		\bottomrule
	\end{tabular}
	\vspace{0.1cm}
    \begin{flushleft}
		{\bf Notes.} Row 1 shows the mean values of \mbh, \lx/\ledd, and \lbol/\ledd\ for NLS1s. Row 2 shows the mean values for BLS1s. Row 3 uses a K--S test to give the probability of finding differences this large given the null hypothesis that NLS1s and BLS1s come from the same population. Row 4 gives the same probability, but this time using a Student's t-test. Using reference source \xte, we were able to compare 35 NLS1s and 54 BLS1s. Using reference source \gro, we compared 13 NLS1s and 51 BLS1s. Using reference source \gx, we compared 25 NLS1s and 51 BLS1s.
	\end{flushleft}
	\label{tab:stats}
\end{table*}

\section*{Acknowledgements}

This research has made use of data obtained through the High Energy Astrophysics Science Archive Research Center Online Service, provided by the NASA/Goddard Space Flight Center. We also thank the anonymous referee for helpful and constructive comments that improved the clarity of the paper.











\bsp	
\label{lastpage}
\end{document}